\documentclass[twocolumn]{aastex62}
\usepackage{amsmath}
\usepackage{graphicx}
\usepackage[normalem]{ulem}

\graphicspath{{./}{figs/}}

\newcommand\redout{\bgroup\markoverwith
{\textcolor{red}{\rule[.5ex]{2pt}{0.4pt}}}\ULon}
\usepackage{xcolor}
\usepackage{threeparttable}
\newcommand{\oii}{[O\,{\sc ii}]}

\newcommand{\mkms}{{\rm km \, s^{-1}}}

\def\cm#1{\, {\rm cm^{#1}}}
\newcommand{\ngold}{five}
\newcommand{\mrvir}{r_{vir}}
\newcommand{\rvir}{$\mrvir$}
\newcommand{\rhophys}{$\rho_{\rm Phys}$}

\newcommand{\rafrb}{$\rm 22h16m4.77s$} % Taken from Cherie's localization (03-16-2020)
\newcommand{\decfrb}{$\rm - 07^\circ53'53.7''$} % Taken from Cherie's localization (03-16-2020)
\newcommand{\zfrb}{0.11778}
\newcommand{\vdmfrb}{339.8}

\newcommand{\hgjname}{J221604.90$-$075356.0}
\newcommand{\hgname}{HG~190608}
\newcommand{\vdmhost}{82}
\newcommand{\vdmhosterr}{35}

\newcommand{\mzzfrb}{z_{\rm host}}
\newcommand{\zzfrb}{$\mzzfrb$}

\newcommand{\mdmunits}{{\rm pc \, cm^{-3}}} 
\newcommand{\dmunits}{$\mdmunits$}
\newcommand{\mdmcosmic}{{\rm DM}_{\rm cosmic}}
\newcommand{\dmcosmic}{$\mdmcosmic$}
\newcommand{\mdmacosmic}{\langle {\rm DM}_{\rm cosmic} \rangle}
\newcommand{\mdmahalos}{\langle {\rm DM}_{\rm halos} \rangle}
\newcommand{\mdmaigm}{\langle {\rm DM}_{\rm IGM} \rangle}
\newcommand{\dmacosmic}{$\mdmacosmic$}
\newcommand{\dmahalos}{$\mdmahalos$}
\newcommand{\dmaigm}{$\mdmaigm$}
\newcommand{\vdmaigm}{54}
\newcommand{\mdmfrb}{{\rm DM}_{\rm FRB}}
\newcommand{\dmfrb}{$\mdmfrb$}
\newcommand{\mdmigm}{{\rm DM}_{\rm IGM}}
\newcommand{\dmigm}{$\mdmigm$}
\newcommand{\mdmhalos}{{\rm DM}_{\rm halos}}
\newcommand{\dmhalos}{$\mdmhalos$}
\newcommand{\mdmhost}{{\rm DM}_{\rm host}}
\newcommand{\dmhost}{$\mdmhost$}
\newcommand{\mdmmwhalo}{{\rm DM}_{\rm MW,halo}}
\newcommand{\dmmwhalo}{$\mdmmwhalo$}
\newcommand{\mdmmwism}{{\rm DM}_{\rm MW,ISM}}
\newcommand{\dmmwism}{$\mdmmwism$}

\newcommand{\mdmfrbc}{{\rm DM}_{\rm FRB,C}}
\newcommand{\dmfrbc}{$\mdmfrbc$}
\newcommand{\mdmslime}{\mdmigm^{\rm slime}}
\newcommand{\dmslime}{$\mdmslime$}

\newcommand{\mrmunits}{\rm rad~m^{-2}}
\newcommand{\rmunits}{$\mrmunits$}

\newcommand{\mmstar}{M_\star}
\newcommand{\mstar}{$\mmstar$}

\newcommand{\mmsun}{{\rm M}_\odot}

\newcommand{\mmhalo}{M_{\rm halo}}
\newcommand{\mhalo}{$\mmhalo$}

\newcommand{\rhoczero}{\rho_{c,0}}  % error Ob h_70^2 from Cooke+18
\newcommand{\thot}{$10^6$ K }

\newcommand{\nrand}{1000} % Number of random fields
\newcommand{\bandwidth}{0.005} % Bandwidth of Mean-Shift clustering
\newcommand{\rhomax}{5~Mpc} %Max impact parameter of search
\newcommand{\mrmax}{r_{\rm max}}
\newcommand{\rmax}{$\mrmax$}
\newcommand{\mfhot}{f_{\rm hot}}
\newcommand{\fhot}{$\mfhot$}

\submitjournal{ApJ}

\shorttitle{FRB 190608 foreground analysis}
\shortauthors{Simha et al.}
\begin{document}

\title{Disentangling the Cosmic Web Towards FRB 190608}

\correspondingauthor{Sunil Simha}
\email{shassans@ucsc.edu}

\author{Sunil Simha}
\affil{University of California - Santa Cruz
1156 High St.
Santa Cruz, CA, USA 95064}

\author{Joseph N. Burchett}
\affil{University of California - Santa Cruz
1156 High St.
Santa Cruz, CA, USA 95064}
\affil{New Mexico State University,
PO Box 30001, MSC 4500,
Las Cruces, NM 88001
}

\author{J. Xavier Prochaska}
\affil{University of California - Santa Cruz
1156 High St.
Santa Cruz, CA, USA 95064}
\affil{Kavli Institute for the Physics and Mathematics of the Universe,
5-1-5 Kashiwanoha, Kashiwa, 277-8583, Japan}

\author{Jay S. Chittidi}
\affil{Maria Mitchell Observatory, 4 Vestal Street, Nantucket, MA 02554, USA}

\author{Oskar Elek}
\affil{University of California - Santa Cruz
1156 High St.
Santa Cruz, CA, USA 95064}

\author{Nicolas Tejos}
\affil{Instituto de F\'isica, Pontificia Universidad Cat\'olica de Valpara\'iso, Casilla 4059, Valpara\'iso, Chile}

\author{Regina Jorgenson}
\affil{Maria Mitchell Observatory, 4 Vestal Street, Nantucket, MA 02554, USA}

\author{Keith W. Bannister}
\affil{Australia Telescope National Facility, CSIRO Astronomy and Space Science, PO Box 76, Epping, NSW 1710, Australia}

\author{Shivani Bhandari}
\affil{Australia Telescope National Facility, CSIRO Astronomy and Space Science, PO Box 76, Epping, NSW 1710, Australia}

\author{Cherie K. Day}
\affil{Australia Telescope National Facility, CSIRO Astronomy and Space Science, PO Box 76, Epping, NSW 1710, Australia}
\affil{Centre for Astrophysics and Supercomputing, Swinburne University of Technology, Hawthorn, VIC 3122, Australia}

\author{Adam T. Deller}
\affil{Centre for Astrophysics and Supercomputing, Swinburne University of Technology, Hawthorn, VIC 3122, Australia}

\author{Angus G. Forbes}
\affil{University of California - Santa Cruz
1156 High St.
Santa Cruz, CA, USA 95064}

\author{Jean-Pierre Macquart}
\affil{International Centre for Radio Astronomy Research, Curtin University, Bentley WA 6102, Australia}

\author{Stuart D. Ryder}
\affil{Department of Physics \& Astronomy, Macquarie University, NSW 2109, Australia}
\affil{Macquarie University Research Centre for Astronomy, Astrophysics \& Astrophotonics, Sydney, NSW 2109, Australia}

\author{Ryan M. Shannon}
\affil{Centre for Astrophysics and Supercomputing, Swinburne University of Technology, Hawthorn, VIC 3122, Australia}



\begin{abstract}
The Fast Radio Burst (FRB) 190608 was detected by the Australian Square-Kilometer Array Pathfinder (ASKAP) and localized to a spiral galaxy at $\mzzfrb=\zfrb$ in the Sloan Digital Sky Survey (SDSS) footprint. The burst has a large dispersion measure ($\rm \mdmfrb=\vdmfrb~\mdmunits$) compared to the expected cosmic average at its redshift. It also has a large rotation measure ($\rm RM_{FRB}=353~ \mrmunits$) and scattering timescale ($\tau=3.3$ ms at 1.28 GHz). \citet{chittidi+20} perform a detailed analysis of the ultraviolet and optical emission of the host galaxy and estimate the host DM contribution to be $110\pm 37~\mdmunits$. This work complements theirs and reports the analysis of the optical data of galaxies in the foreground of FRB 190608 in order to explore their contributions to the FRB signal. Together, the two studies delineate an observationally driven, end-to-end study of matter distribution along an FRB sightline; the first study of its kind. Combining our Keck Cosmic Web Imager (KCWI) observations and public SDSS data, we estimate the expected cosmic dispersion measure
\dmcosmic\ along the sightline to FRB 190608. We first estimate 
the contribution of hot, ionized gas in intervening virialized halos ($\mdmhalos \approx 7-28~\mdmunits$). 
Then, using the Monte Carlo Physarum Machine (MCPM) methodology, we produce a 3D map of ionized gas in cosmic web filaments and compute the DM contribution from matter outside halos ($\mdmigm \approx 91-126~\mdmunits$). 
This implies a greater fraction of ionized gas along this sightline is extant outside virialized halos. We also investigate whether the intervening halos can account for the large FRB rotation measure and pulse width and conclude that it is implausible. Both the pulse broadening and the large Faraday rotation likely arise from the progenitor environment or the host galaxy.
\end{abstract}


\section{Introduction}
Galaxies are the result of gravitational accretion of baryons onto dark
matter halos, i.e. the dense gas that has cooled and condensed to 
form dust, stars, and planets. The dark matter halos, according to simulations, are embedded in the cosmic web, a filamentous structure of matter \citep[e.g.][]{Springel+05}. The accretion process of galaxies is 
further predicted, at least for halo masses $\mmhalo \gtrsim10^{12}~\mmsun$, to generate a halo of baryons, most likely
dominated by gas shock-heated to the virial temperature of
    the potential well \citep{White1978, White1991,Kauffmann1993,Somerville1999, Cole2000}.
At $T \gtrsim$ \thot  and $n_e \sim 10^{-4} \cm{-3}$, however,
this halo gas is very difficult to detect in emission \citep{Kuntz2000,Yoshino+2009,Henley2013}
and similarly challenging to observe in absorption \citep[e.g.][]{Burchett+19}.
And while experiments leveraging the Sunyaev-Zeldovich effect
are promising \citep{planck_sz}, these are currently limited to
massive halos
and are subject
to significant systematic effects \citep{Lim+20}.

Therefore, there has been a wide range of predictions
for the mass fraction of baryons in massive halos that range from 
$\approx 10\%$ to nearly the full complement relative to the
cosmic mean $\Omega_b / \Omega_m$ \citep{tng1}. Here, $\Omega_b$ and $\Omega_m$ are the average cosmic densities of baryons and matter respectively.
Underlying this order-of-magnitude spread in predictions
are uncertain physical processes that eject gas from galaxies and
can greatly shape them and their environments \citep[e.g.][]{Suresh+15}.

Fast radio bursts (FRBs) are dispersed by intervening ionized matter such that the pulse arrival delay, with respect to a reference frequency, scales as the inverse square of frequency times the DM.
The DM is the path integral of the
electron density, $n_e$, weighted by the scale
factor $(1+z)^{-1}$, i.e. ${\rm DM} \equiv \int n_e \, ds / (1+z)$. These FRB measurements are sensitive to all of the ionized
gas along the sightline. Therefore, they have the potential
to trace the otherwise invisible plasma surrounding and
in-between galaxy halos~\citep{jpm+2020}. The Fast and Fortunate for FRB Follow-up (F$^4$)
team\footnote[1]{http://www.ucolick.org/f-4} has initiated a program to disentangle the cosmic web
by correlating the dispersion measure (DM) of 
fast radio bursts (FRBs) with the distributions of 
foreground galaxy halos \citep{mcquinn14,xyz19}.
This manuscript marks our first effort.

Since the DM is an additive quantity, it may be split into individual 
contributions of intervening, ionized gas reservoirs:
\begin{equation}
    \begin{aligned}
        \mdmfrb =~\rm DM_{MW}+ \mdmcosmic + \mdmhost
    \end{aligned}
    \label{eqn:dmbudget}
\end{equation}
Here, $\rm DM_{\rm MW}$ refers to the contribution from the Milky Way
which is further split into its ISM and halo gas contributions
(\dmmwism\ and  \dmmwhalo\ respectively). 
Additionally, \dmhost\ is the net contribution from the host galaxy and its halo, including any contribution from the immediate environment of 
the FRB progenitor. 
Meanwhile, \dmcosmic\ is the sum of contributions from gas in the circumgalactic medium (CGM) of intervening halos (\dmhalos)\  
and the intergalactic medium (IGM; \dmigm). 
Here, CGM refers to the gas found within dark matter halos including the intracluster medium of galaxy clusters, 
and the IGM refers to gas between galaxy halos.

\cite{jpm+2020} have demonstrated that the FRB population
defines a cosmic DM-$z$ relation that closely tracks the
prediction of modern cosmology \citep{inoue04,xyz19,DengZhang14},
i.e., the average cosmic DM is
\begin{equation}
    \mdmacosmic = \int\limits_0^{\mzzfrb} \bar n_e(z)\frac{cdz}{H(z)(1+z)^2}  
    \label{eqn:DMzRelation}
\end{equation}
with  $\bar n_e = f_d(z) \rho_b(z)/m_p (1-Y_{\rm He}/2)$, which is the mean density of electrons at redshift $z$. Here, $m_p$ is the proton mass, $Y_{\rm He} = 0.25$ is the mass fraction of Helium (assumed doubly ionized in this gas),
$f_d(z)$ is the fraction of cosmic baryons in diffuse ionized gas,
i.e. excluding dense baryonic phases 
such as stars and neutral gas \citep[see ][ and Appendix \ref{appendix:f_d}]{jpm+2020}.
$\rho_b(z) = \Omega_{b,0} \rhoczero (1+z)^3$,
$\rhoczero$ is the critical density at $z=0$,
and $\Omega_{b,0}$ is the baryon energy density 
today relative to $\rhoczero$. $c$ is the speed of light in vacuum and $H(z)$ is the Hubble parameter. Immediately relevant to the study at hand, for FRB 190608, $\mdmacosmic \approx 100$ \dmunits\ at \zzfrb = \zfrb.

\begin{figure}[h]
    \includegraphics[width=\linewidth]{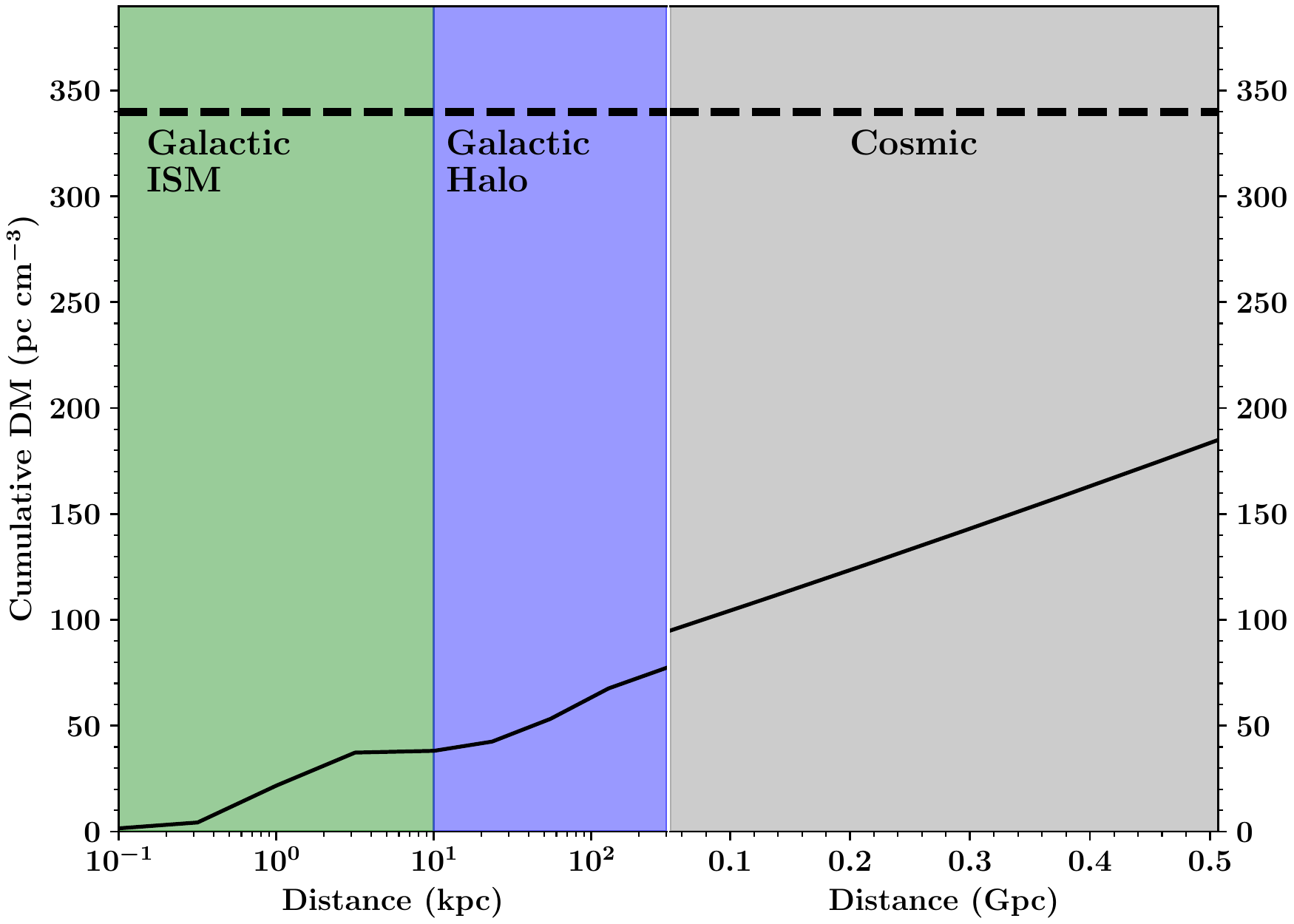}
    \caption{\footnotesize\textbf{The cumulative FRB dispersion measure for FRB~190608}. 
    The dashed line corresponds to the $\mdmfrb = \vdmfrb \, \mdmunits$
    reported for the FRB \citep{day+2020}, which is at the highest
    distance shown ($\approx 0.5$\,Gpc).
    The solid curve is an estimate of the cumulative DM
    moving out from Earth towards the FRB. 
    The Milky Way's ISM \citep[green; model of][]{ne2001} 
    and halo \citep[blue; model of][]{xyz19} together may contribute $\approx 100~\mdmunits$. 
    If the foreground cosmic web (grey) contributes the expected average 
    (Equation~\ref{eqn:DMzRelation}), this adds an additional $\approx 100~\mdmunits$
    as modeled. Note that the horizontal axis is discontinuous at the Halo-Cosmic interface and this is the reason for a discontinuous cumulative DM.
    The difference between the solid and dashed lines at the FRB
    is $\approx 160~\mdmunits$ and is expected to be attributed to 
    the host galaxy and/or an above average contribution from the cosmic web
    (e.g.\ overdensities in the host galaxy foreground). 
    }
    \label{fig:dm_cumul}
\end{figure}

Of the \ngold\ FRBs in the \cite{jpm+2020} `gold' sample,
FRB~190608 exhibits a \dmcosmic~value well in excess
of the average estimate for its redshift:
${\rm DM_{\rm cosmic}} / \mdmacosmic \approx 2$
based on the estimated contributions of \dmmwhalo\
and \dmhost.
This is illustrated in Fig.~\ref{fig:dm_cumul},
which compares the measured $\mdmfrb = \vdmfrb \, \mdmunits$ \citep{day+2020}
with the cumulative contributions from the Galactic ISM \citep[taken as \dmmwism = 38~\dmunits;][]{ne2001},
the Galactic halo \citep[taken as \dmmwhalo = 40~\dmunits;][]{xyz19}, and the average cosmic web 
(Equation~\ref{eqn:DMzRelation}). 
These fall $\approx 160 \, \mdmunits$ short of the observed value.
\citet{chittidi+20} estimate the
host galaxy ISM contributes
$\rm DM_{host,ISM}=\vdmhost \pm \vdmhosterr~\mdmunits$ based on the observed
H$\beta$ emission measure and $\rm DM_{host,halo}=28 \pm 13\, \mdmunits$ for the host
galaxy's halo, thus nearly accounting for the deficit. The net \dmhost~is therefore taken here to be $110\pm37~\mdmunits$.

While these estimates almost fully account for the large \dmfrb, several of them bear significant
uncertainties (e.g., \dmmwhalo \, and \dmhost).
Furthermore, we have assumed the average \dmcosmic\ value, a quantity
predicted to exhibit significant variance from sightline
to sightline \citep{mcquinn14,xyz19,jpm+2020}. Therefore, in this work  we examine the galaxies and large-scale structure foreground to FRB~190608 
to analyze whether $\mdmcosmic\approx\mdmacosmic$ or whether there 
is significant deviation from the cosmic average.
These analyses constrain several theoretical expectations
related to \dmacosmic\ \citep[e.g.][]{mcquinn14,xyz19}. 
In addition, FRB 190608 exhibits a relatively large rotation measure ($\rm RM = 353~\mrmunits$) and a large, frequency dependent exponential tail ($\tau_{1.4 {\rm Ghz}}=2.9$\,ms) in its temporal pulse profile that corresponds to scatter-broadening \citep{day+2020}. We explore the possibility that these arise from foreground matter overdensities
and/or galactic halos \cite[similar to the analysis by][]{frb181112}.

This paper is organized as follows. In Section~2, we present our data on the host and foreground galaxies and our spectral energy distribution (SED) fitting method for determining galaxy properties. In Section~3, we describe our methods and models in estimating the separate \dmcosmic\  contributions from intervening halos and the diffuse IGM. Section~4 explores the possibility of a foreground structure accounting for the FRB rotation measure and pulse width. Finally, in Section~5, we summarise and discuss our results.
Throughout our analysis, we use cosmological parameters derived from the results of \citet{Planck15}.

%%%%%%%%%%%%%%%%%%%%%%%%%%%%%%%%%%%%%%%%%%%%%%%%%%%%%%%%%%%%%%%%%%%%%%%%%

%%%%%%%%%%%%%%%%%%%%%%%%%%%%%%%%%%%%%%%%%%%%%%%%%%%%%%%%%%%%%%%%%%%
%%%%%%%%%%%%%%%%%%%%%%%%%%%%%%%%%%%%%%%%%%%%%%%%%%%%%%%%%%%%%%%%%%%
\section{Foreground Galaxies}

%%%%%%%%%%%%%%%%%%%%%%%%%%%%%%%%%%%
\subsection{The Dataset}\label{sec:dataset}

FRB 190608 was detected and localized by the Australian Square Kilometre Array Pathfinder (ASKAP) to RA = \rafrb, Dec = \decfrb \citep{day+2020},
placing it in the outer disk of the 
galaxy \hgjname\ at $z=\zfrb$ (hereafter \hgname) 
cataloged by the Sloan Digital Sky Survey (SDSS).

To search for nearby foreground galaxies, we obtained six $33''\times20''$ integral field unit (IFU) exposures (1800 s each) using the Keck Cosmic Web Imager \citep[KCWI;][]{KCWIDRP} in a mosaic centered at the host galaxy centroid. The IFU was used in the ``large" slicer position with the ``BL" grating, resulting in a spectral resolution, $R_0\sim900$. The six exposures cover an approximately $1' \times 1'$ field around the FRB host. They were reduced using the standard KCWI reduction pipeline \citep{KCWIDRP} with sky subtraction
\citep[see][for additional details]{chittidi+20}.

From the reduced cubes, we extracted the spectra of sources identified in the white-light images using the Source Extractor and Photometry (SEP) package \citep{Barbary2016,SExtractor}. We set the detection threshold to 1.5 times the estimated RMS intensity after background subtraction and specified a minimum source area of 10 pixels ($\sim5$ kpc at $z = 0.05$)
to be a valid detection. Thirty sources were identified this way across the six fields; 
none have SDSS spectra. SEP determines the spatial light profiles of the sources and for each source outputs major and minor axis values of a Gaussian fit. Using elliptical apertures with twice those linear dimensions, we extracted source spectra. We then determined their redshifts using the Manual and Automatic Redshifting Software \citep[MARZ,][]{marz}. MARZ fits each spectrum with a template spectrum and determines the redshift corresponding to the maximum cross-correlation. Seven objects had unambiguous redshift estimates, whereas the rest did not show any identifiable line emission. 
Five of the seven objects with secure redshifts are at $z>\mzzfrb$ and are not discussed further. We observed two objects (RA = $22^\mathrm{h}16^\mathrm{m}4.86^\mathrm{s}$, Dec = $-7^\circ53'44.16''$ eq. J2000) with a single strong emission feature at 4407\,$\rm \AA$ for one and 3908 $\rm \AA$ for the other. MARZ reported high cross-correlations with its templates for when this feature was associated with either the \oii 3727-3729 $\rm \AA$ doublet (corresponding to $z < z_{\rm FRB}$) or Ly$\alpha$ (corresponding to $z > 2$). There are no other discernible emission lines in the spectra. If we assume the emission line is indeed \oii, we can then measure the the peak intensity of H$\beta$. Thus, in both spectra, the H$\beta$ peak would be less than 0.02 times the \oii\ peak intensity, which would imply an impossible metallicity. Thus we conclude that the features are likely Ly$\alpha$ and place these as galaxies at $z> 2.6$.

In the remaining 23 spectra, we detect no identifiable emission lines. Since we measure only weak continua (per-pixel $\rm SNR<1$), if any, from the remaining 23 objects, we find it difficult to estimate the likelihood of their being foreground objects from synthetic colors.

We experimented with decreasing the minimum detection area threshold to 5 pixels. This increases the number of detected sources, but the additional sources, assuming they are actually astrophysical, do not have any identifiable emission lines. These sources are most likely fluctuations in the background.

To summarize, we found no foreground galaxy in the 1 arcmin sq.\ KCWI field. Assuming the halo mass function derived from the Aemulus project \citep{HMF}, the average number of foreground halos (i.e., for $z<\mzzfrb$ and in a $1'\times1'$ field) between $2\times10^{10}~\mmsun$ and $10^{16}~\mmsun$ is 0.23; 
therefore, the absence of objects can be attributed to Poisson variance. This general conclusion remains valid even when we refine the expected number of foreground halos based on the inferred overdensities along the line of sight (see Section~\ref{sec:cosmicWeb}).

To expand the sample,
we then queried the SDSS-DR16 database for all spectroscopically
confirmed galaxies with impact parameters $b < 5$\,Mpc
(physical units) to the FRB sightline and $z < \mzzfrb$. This impact parameter threshold was chosen to encompass any galaxy
or large-scale structure that might contribute
to \dmcosmic\ along the FRB sightline. As the FRB location lies in one of the narrow strips in the SDSS footprint, the query is spatially truncated in the north-eastern direction. Effectively no object with $b \gtrsim 2.5~\rm Mpc$ in that direction was present in the query results due to this selection effect.

We further queried the SDSS database for all galaxies with 
photometric redshift estimates such that $z_{\rm phot}-2\delta z_{\rm phot} < \mzzfrb$ and $z_{\rm phot}/\delta z_{\rm phot} >1$. Here $\delta z_{\rm phot}$ is the error in $z_{\rm phot}$ reported in the database. We rejected objects that were flagged as cosmic rays or were suspected cosmic rays or CCD ghosts. None of these recovered galaxies lie within 250~kpc of the sightline as estimated
from $z_{\rm phot}$. However, several galaxies were found with $z_{\rm phot}>\mzzfrb$ and $z_{\rm phot}-2\delta z_{\rm phot}<\mzzfrb$ that can be within 250 kpc if their actual redshifts were closer to $z_{\rm phot}-2\delta z_{\rm phot}$.

%%%%%%%%%%%%%%%%%%%%%%%%%%%%%%%%%%%%%%%%%%%%%%%%%%%%%%%%%%%%%%%%%%%
\subsection{Derived Galaxy Properties}

For each galaxy in the spectroscopic sample, we have 
estimated its stellar mass, \mstar,~by fitting the SDSS \textit{ugriz} photometry with an SED using CIGALE \citep{cigale}. We assumed, for simplicity, a delayed-exponential star-formation history with no burst population, a synthetic stellar population prescribed by \citet{bc03}, the
\citet{Chabrier03} initial mass function (IMF), dust attenuation models
from \citet{calzetti01}, and dust emission templates from \citet{dale14},  where the AGN fraction was capped at 20\%. 
The models typically report a $\lesssim 0.1$~dex statistical uncertainty on
\mstar\ and star formation rate from the SED fitting, but we estimate systematic uncertainties are
$\approx 2 \times$ larger.
Table~\ref{tab:fg_gals} lists the observed and derived properties
for the galaxies.

Central to our estimates of the contribution of 
halos to the DM is an estimate of the halo mass, \mhalo.
A commonly adopted procedure is to estimate \mhalo\ 
from the derived stellar mass, \mstar,\ by using the  abundance matching technique.
Here, we adopt the stellar-to-halo-mass ratio (SHMR) of \cite{moster+13}, which 
also assumes the Chabrier IMF. Estimated halo masses of
the foreground galaxies
range from $10^{11}~\mmsun$ to $\gtrsim 10^{12}~\mmsun$.

\begin{table*}
\centering
\caption{\footnotesize Observed and derived properties of the spectroscopic foreground galaxies from SDSS.\tablenotemark{$\dagger$}\label{tab:fg_gals}}
\begin{tabular}{|c|c|c|c|c|c|c|c|c|c|c|}
\hline
RA & Dec & u & g & r & i & z & Redshift & $b$ & $\log(M_*/M_\odot)$ & $\log(M_{halo}/M_\odot)$ \\
$\deg$ & $\deg$ & mag & mag & mag & mag & mag &  & kpc &  &  \\
\hline
334.00914 & -7.87554 & 18.73 & 17.54 & 16.98 & 16.63 & 16.37 & 0.09122 & 158 & 10.36 & 11.81 \\
333.97368 & -7.87678 & 19.28 & 17.87 & 16.95 & 16.50 & 16.20 & 0.08544 & 300 & 10.59 & 12.09 \\
333.88476 & -8.01812 & 18.48 & 17.39 & 16.92 & 16.72 & 16.59 & 0.02732 & 367 & 9.06 & 11.04 \\
334.01930 & -8.02294 & 18.31 & 16.58 & 15.74 & 15.38 & 15.13 & 0.06038 & 541 & 10.63 & 12.17 \\
334.04856 & -7.79251 & 19.89 & 17.99 & 17.05 & 16.63 & 16.19 & 0.07745 & 597 & 10.54 & 12.01 \\
333.77207 & -7.53690 & 19.79 & 17.99 & 17.51 & 17.24 & 17.01 & 0.02394 & 784 & 8.85 & 10.95 \\
334.07667 & -7.76554 & 19.43 & 18.24 & 17.39 & 16.96 & 16.61 & 0.08110 & 819 & 10.37 & 11.82 \\
333.99058 & -8.10044 & 19.97 & 18.48 & 17.88 & 17.47 & 17.30 & 0.06522 & 951 & 9.75 & 11.39 \\
334.08866 & -8.01256 & 18.95 & 18.38 & 17.29 & 16.84 & 16.56 & 0.11726 & 1050 & 10.91 & 12.79 \\
334.12864 & -8.08630 & 19.20 & 17.86 & 17.31 & 16.96 & 16.75 & 0.07096 & 1091 & 10.01 & 11.55 \\
\hline
\end{tabular}
\tablenotetext{\dagger}{\footnotesize This table is published in its entirety in the machine-readable format. Ten galaxies with the lowest impact parameters are shown here.}
\end{table*}

\begin{figure}[t!]
    \centering
    \includegraphics[width=\linewidth]{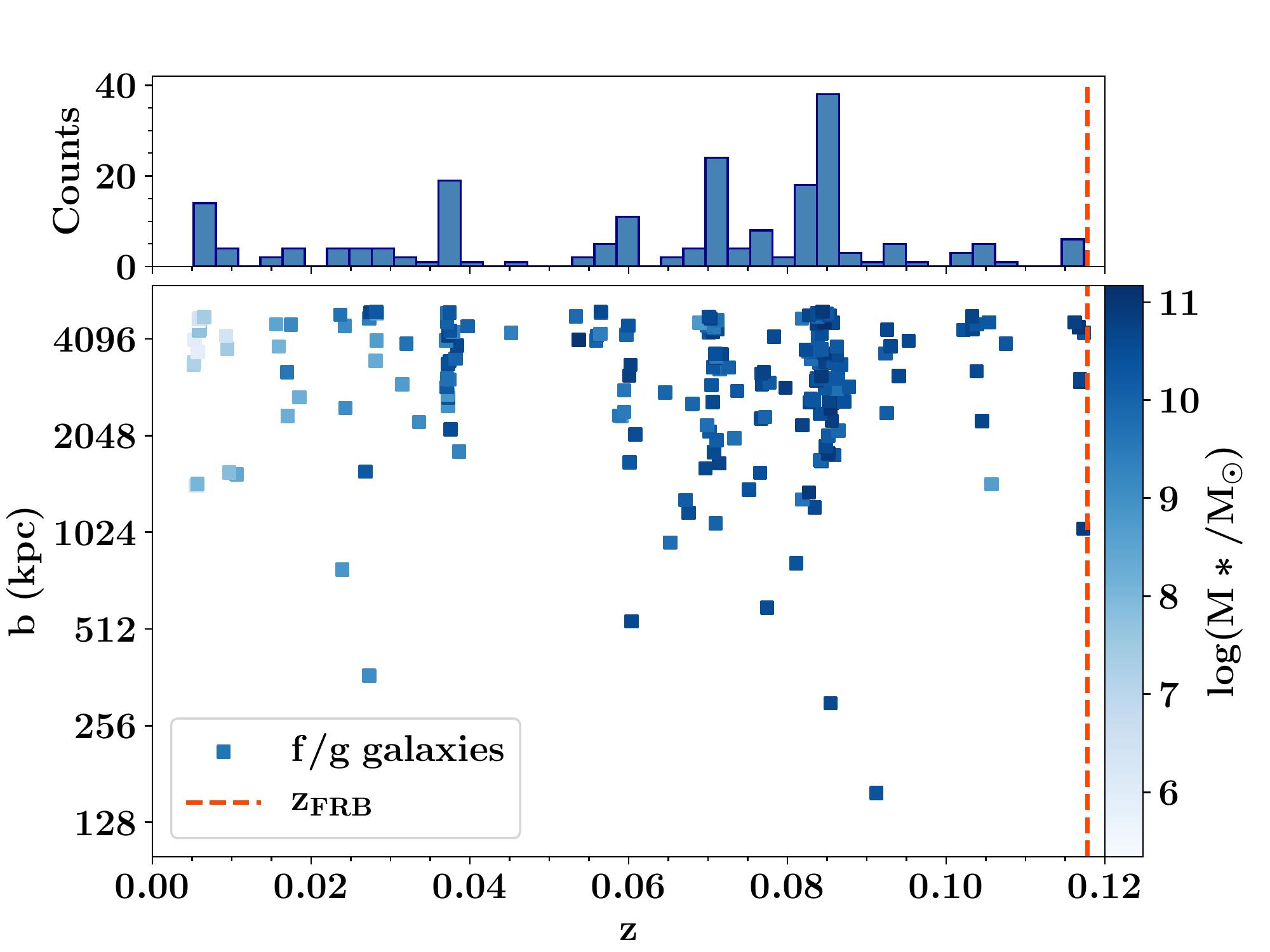}
    \caption{\footnotesize\textbf{The spatial distribution of foreground galaxies}.\textit{(Bottom)} A scatter plot of foreground galaxy redshifts, $z$, and impact parameters, $b$. The points are colored according to the estimated stellar masses. The red 
    dashed-line indicates the FRB host redshift. \textit{(Top)} A histogram of the redshifts. The `spikes' in the distribution, e.g.\ 
    at $z \sim 0.08$, indicate overdensities in the 
    underlying cosmic web structure.}
    \label{fig:rho_z}
\end{figure}

%%%%%%%%%%%%%%%%%%%%%%
\subsection{Redshift distribution of foreground galaxies}

Fig.~\ref{fig:rho_z} shows the distribution of impact parameters and spectroscopic redshifts for the foreground galaxies.
There is a clear excess of galaxies at $z \sim 0.08$.
Empirically, there are 50~galaxies within a redshift range $\Delta z=0.005$ of $z = 0.0845$.
A review of group and cluster catalogs of the SDSS \citep{Yang+07, Rykoff+14},
however, shows no massive 
collapsed structure ($\mmhalo >10^{13} \mmsun$) 
at this redshift and within $b = 2.5$\,Mpc of the sightline.  
The closest redMaPPer cluster at this redshift is at a transverse distance of 8.7~Mpc. However, we must keep in mind that the survey is spatially truncated in the north-eastern direction and therefore we cannot conclusively rule out the presence of a nearby galaxy group or cluster. Nevertheless,
the distribution suggests an overdensity of galaxies tracing some form of large-scale
structure, e.g.\ a filament connecting this distant cluster to another (see Section \ref{sec:cosmicWeb}).

To empirically assess the statistical significance of 
FRB~190608 exhibiting an excess of foreground galaxies
(which would suggest an excess \dmcosmic),
we performed the following analysis.  First, we defined a 
grouping\footnote{We avoid the use of group or cluster to minimize
confusion with those oft used terms in astronomy.} of galaxies
using a Mean-Shift clustering algorithm on the galaxy
redshifts in the field adopting a bandwidth $\rm \Delta z$ of \bandwidth\
($\approx 3100~ \mkms$). 
This generates a redshift centroid and the number of galaxies in a series
of groupings for the field.
For the apparent overdensity, we recover $z = 0.0843$ and $N=62$ galaxies; this is the grouping with the highest 
cardinality in the field.
We then generated \nrand\ random sightlines in the SDSS footprint
and obtained the redshifts of galaxies with 
$z < \mzzfrb$ and with impact parameters $b < \rm \rhomax$, 
restricting the sample to galaxies with $z> 0.02$ for computational expediency. We also restricted the stellar masses to lie above $10^{9.3}~\mmsun$ to account for survey completeness near z = 0.08.
This provides a control sample for comparison with the FRB~190608 field.

\begin{figure}
    \centering
    \includegraphics[width=\linewidth]{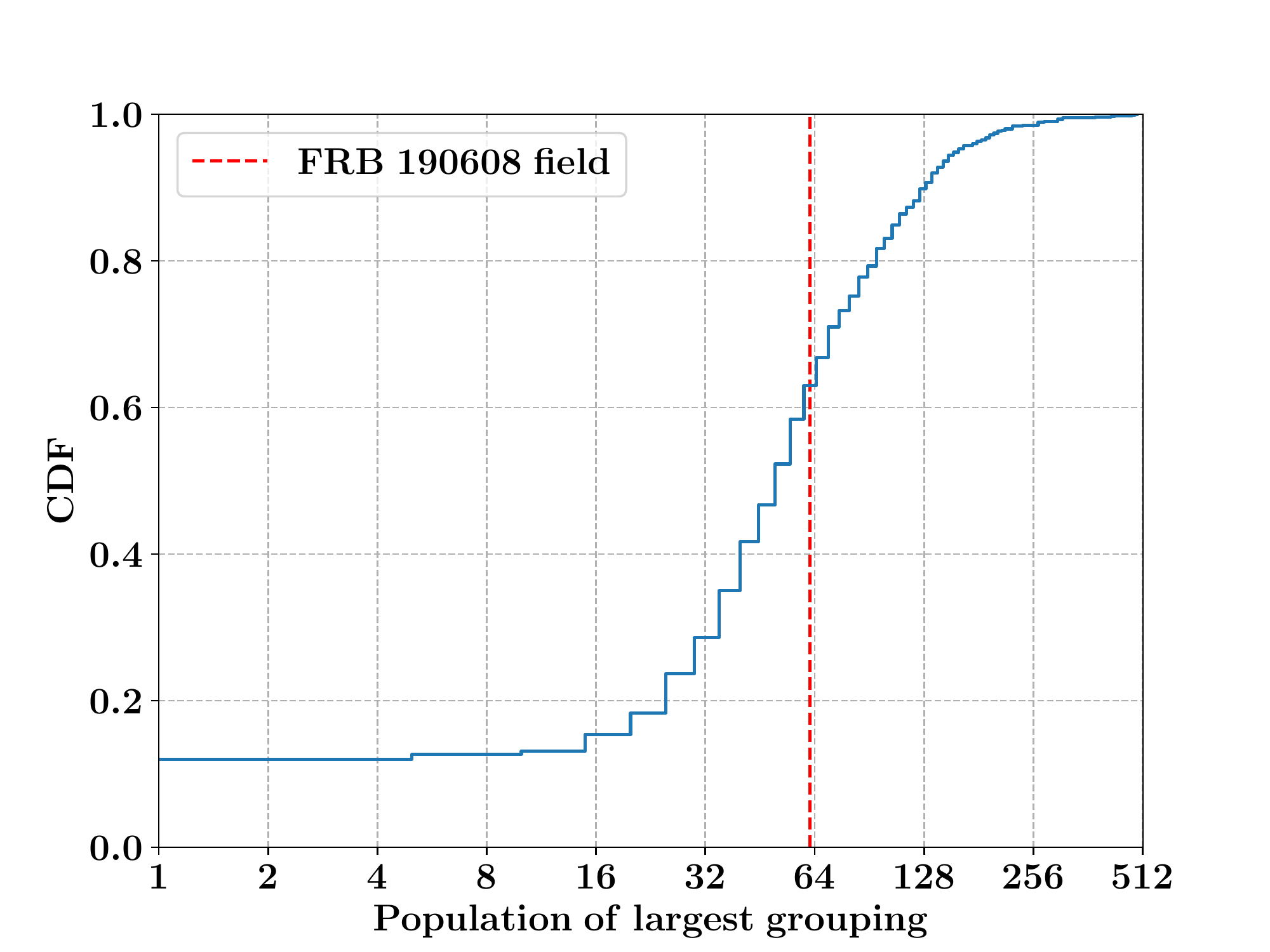}
    \caption{\footnotesize\textbf{Grouping population sizes in SDSS fields}. A cumulative histogram of the sizes of the most populous redshift groupings in \nrand~random SDSS fields. Each field was searched for galaxies more massive than $10^{9.3}~\mmsun$ with spectra within \rhomax\ of a sightline passing through the center. The groupings are computed using a Mean-Shift algorithm with bandwidth $\Delta z =$\bandwidth. Their centroids all lie between z = 0.02 and \zzfrb. The most populous redshift grouping found in the FRB field at $z\sim 0.08$ is indicated by the dashed, red line. {At the $63^{\mathrm{rd}}$ percentile, the FRB field does not have rare overdensities in its foreground.}}
    \label{fig:cluster}
\end{figure}

Fig.~\ref{fig:cluster} shows the cumulative distribution 
of the number of galaxies in the most populous groupings in each field.
We find that the FRB field's largest grouping is at the $63^{\mathrm{rd}}$ percentile, and therefore conclude that it is not a 
rare overdensity.  It might, however, make a significant contribution to \dmcosmic, a hypothesis that we explore in the next section.
%%%%%%%%%%%%%%%%%%%%%%%%%%%%%%%%%%%%%%%%%%%%%%%%%%%%%%%%%
%%%%%%%%%%%%%%%%%%%%%%%%%%%%%%%%%%%%%%%%%%%%%%%%%%%%%%%%%
\section{DM Contributions} %Host and Foreground Halos}
\label{sec:DM}

This section estimates \dmhalos, and \dmigm. For the sake of clarity, we make a distinction in the terminology we use to refer to the cosmic contribution to the dispersion measure estimated in two different ways. First, we name  the difference between \dmfrb\ and the estimated host and Milky Way contributions \dmfrbc\ i.e. $\mdmfrbc = \mdmfrb -\rm DM_{MW} -  \mdmhost \approx 152 \, \mdmunits$. Second, we shall henceforth use the term \dmcosmic~ to refer to the sum of \dmhalos~and \dmigm~semi-empirically estimated from the foreground galaxies.

\subsection{Foreground halo contribution to \dmcosmic}
\label{sec:fgHalos}
\begin{figure*}[th!]
    \centering
    \includegraphics[width=\linewidth]{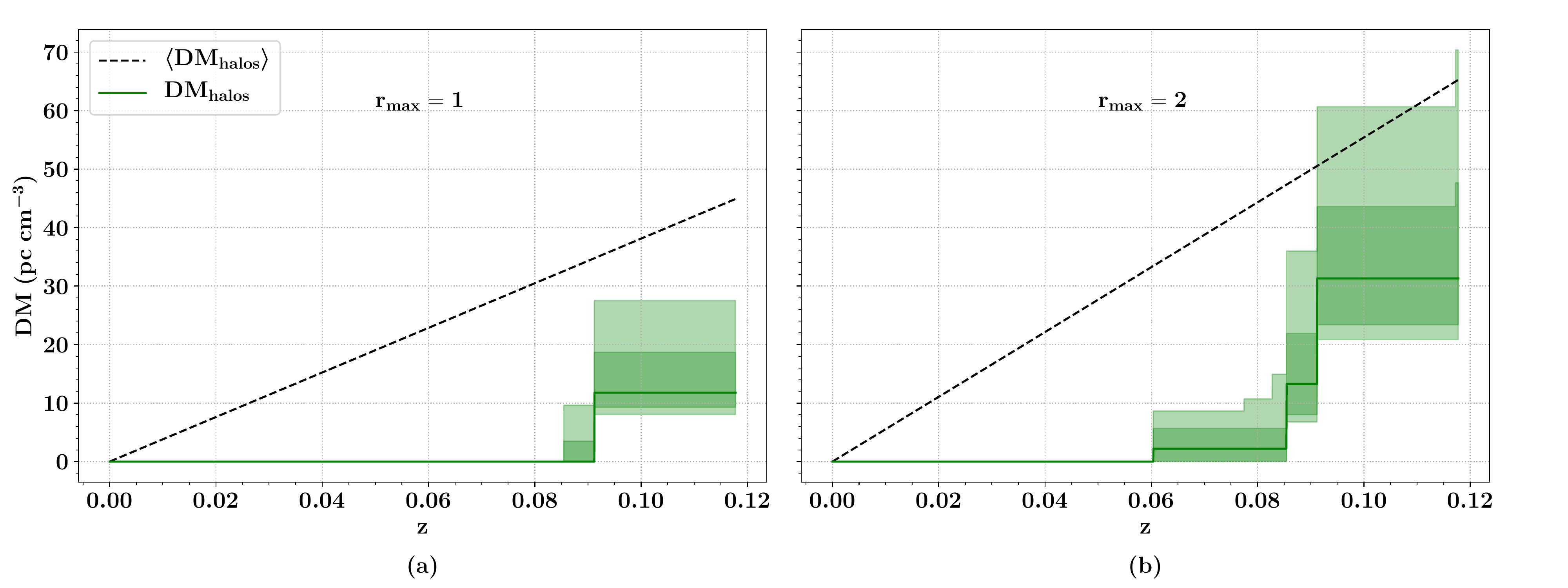}
    %    {\bf Figure \figfrb:} 
    \caption{\footnotesize \textbf{\dmhalos~vs redshift}. The black line represents \dmahalos,\ i.e., the average DM from halos using the Aemulus halo mass function (ignoring the IGM). The solid green line is our estimate of \dmhalos, the DM contribution from intervening halos of galaxies found in SDSS and assuming a hot gas fraction 
    $\mfhot = 0.75$. The dark green shaded region is obtained by varying the stellar masses of each of the intervening halos by 0.1 dex, which modulates the adopted halo mass. This is representative of the uncertainty in DM propagated from stellar mass estimation. The lighter green shaded region is obtained by similarly varying the stellar masses by 0.16 dex and it is representative of the uncertainty in DM propagated from the scatter in the SHMR.
    This calculation was performed for two values of the dimensionless radial extent of the halo's matter distribution, $r_{max}$: 1 (\textit{left}) and 2 (\textit{right}). Using the central measures of stellar mass and the SHMR, the intervening galaxies contribute \dmhalos\ less than the expected cosmic average, \dmahalos, and do not exceed 50 \dmunits.
    }
    \label{fig:dm_z_sdss}
\end{figure*}

We first consider the DM contribution from
halo gas surrounding foreground galaxies, \dmhalos.
For the four galaxies with $b < 550$\,kpc, 
all have estimated halo masses 
$\mmhalo \leq 10^{12.2}~\mmsun$.  We adopt the definition of \rvir~using the formula for average virial density from \citet{BryanNorman98}, i.e. the average halo density enclosed within \rvir~is:

\begin{equation}
    \begin{aligned}
    \rho_{vir} &=(18\pi^2-82q-39q^2)\rho_{c}\\
    q &= \frac{\Omega_{\Lambda,0}}{\Omega_{m,0}(1+z)^3+\Omega_{\Lambda,0}}
    \end{aligned}
\end{equation}
Here $\rho_c$ is the critical density of the universe at redshift $z$ and $\Omega_{\Lambda,0}$ is the dark energy density relative to $\rho_{c,0}$. Computing \rvir~from the estimated halo masses we find that only the halo with the smallest impact parameter at $z=0.09122$ (i.e. first entry in Table \ref{tab:fg_gals}) is intersected by
the sightline.  In the following, however, 
we will allow for uncertainties in \mhalo\ and also
consider gas out to 2\rvir.
Nevertheless, we proceed with the expectation 
that \dmhalos\ is small.

To derive the DM contribution from each halo, we must
adopt a gas density profile and the total mass of baryons
in the halo.  For the former,
we assume a modified Navarro-Frenk-White (NFW) baryon profile as described in \citet{xyz19},
with profile parameters $\alpha = 2$ and $y_0 = 2$. 
We terminate the profile at a radius \rmax, given in units of \rvir\
(i.e.,\ \rmax=1 corresponds to \rvir). 
The gas composition is assumed to be primordial,
i.e., 75\% hydrogen and 25\% helium by mass. 
For the halo gas mass, we define 
$M^b_{\rm halo} \equiv \mfhot (\Omega_b/\Omega_m) \mmhalo$,
with \fhot\ parametrizing the fraction of the total baryonic
budget present within the halo as hot gas.
For a halo that has effectively retained all of its baryons,
a canonical value is $\mfhot = 0.75$, which allows for 
$\approx 25\%$ of the baryons to reside in stars, remnants, and neutral
gas of the galaxy at its center \citep[e.g. ][]{Fukugita+98}.
If feedback processes have effectively removed gas from the halo,
then $\mfhot \ll 0.75$. For simplicity, we do not vary \fhot~with halo mass but this fraction might well be a function of halo properties \citep[e.g. ][]{Behroozi+10}.

At present, we have only weak constraints on 
\fhot, $\alpha$, and $y_0$, and we emphasize that 
our fiducial values
are likely to maximize the DM estimate for a given 
halo (unless the impact parameter is $\ll \mrvir$).
We therefore consider the estimated \dmhalos\ to be an upper bound.
However, we further note that the choice of \rmax,\ 
which effectively sets the size of the gaseous halo
is largely arbitrary.  In the following, we consider
$\mrmax = 1$ and 2.

The DM contribution of each foreground halo was computed by estimating the column density of free electrons intersecting 
the FRB sightline.
Fig.~\ref{fig:dm_z_sdss}a shows the estimate of \dmhalos\ for \rmax~= 1. When \rmax~= 2 (Fig.~\ref{fig:dm_z_sdss}b), the halo at $z=0.09122$ (Table \ref{tab:fg_gals}) contributes an additional $\sim10~\mdmunits$ to the \dmhalos~estimate from the extended profile. Furthermore, the halo at $z=0.08544$ contributes $\sim10~\mdmunits$ and the halo at $z=0.06038$ contributes $\sim2~\mdmunits$.

\begin{table*}
\centering
\caption{\footnotesize SDSS galaxies with photometric redshifts that potentially contribute to \dmhalos\label{tab:phot_z}}
\begin{tabular}{|c|c|c|c|c|c|c|c|c|c|}
\hline
RA & Dec & Separation from FRB & u & g & r & i & z & $z_{phot}$ & $\delta z_{phot}$ \\
$\deg$ & $\deg$ & arcmin & mag & mag & mag & mag & mag &  &  \\
\hline
334.01251 & -7.88616 & 0.84 & 22.09 & 20.41 & 19.56 & 19.34 & 18.94 & 0.21 & 0.06 \\
334.03281 & -7.90426 & 0.86 & 23.62 & 22.18 & 21.25 & 20.94 & 20.22 & 0.27 & 0.12 \\
334.03590 & -7.88558 & 1.22 & 22.89 & 22.33 & 21.31 & 21.08 & 19.63 & 0.34 & 0.13 \\
334.00943 & -7.87979 & 1.26 & 21.00 & 20.04 & 19.40 & 19.16 & 18.95 & 0.15 & 0.04 \\
\hline
\end{tabular}
\end{table*}

In addition to the spectroscopic sample, we performed a similar analysis on the sample of galaxies with $z_{\rm phot}$ only. As mentioned earlier, no galaxy in this sample was found within 250 kpc if their redshift was assumed to be $z_{\rm phot}$ and therefore,  their estimated contribution to \dmhalos~was null. However, if we assumed their redshifts were $z_{\rm phot}-2\delta z_{\rm phot}$, we estimate a net DM contribution of $\sim30~\mdmunits$ from four galaxies (Table \ref{tab:phot_z}). Their contribution decreases with increasing assumed redshift.  At \zzfrb, only the first two galaxies contribute and their net contribution is estimated to be $\sim13~\mdmunits$. A spectroscopic follow-up is necessary to pin down the galaxies' redshifts and therefore their DM contribution as they lie outside our the field of view of our KCWI data.

Using the aforementioned assumptions for the halo gas profile, we can compute the average contribution to \dmacosmic, i.e. \dmahalos, by estimating the  fraction of cosmic electrons enclosed in halos, $f_{\rm e,halos}(z)$. \dmahalos~provides a benchmark that we may compare against \dmhalos. First, we find the average density of baryons found in halos between $10^{10.3}~\mmsun$ and $10^{16}~\mmsun$ using the Aemulus halo mass function \citep{HMF}, i.e. $\rho_{b,\rm halos}(z)$. The ratio of this density to the cosmic matter density $\rho_{b}(z)$ is termed $f_{\rm halos}$. Then, according to our halo gas model, $f_{\rm e,halos}(z)$ is:
\begin{equation}
    \begin{aligned}
        f_{\rm e,halos}(z) &= \frac{\bar{n}_{e,\rm halos}(z)}{\bar{n}_e(z)} = \frac{\rho_{b,\rm halos}(z)\mfhot}{\rho_b(z)f_d(z)}\\
        &=f_{\rm halos}(z)\frac{\mfhot}{f_d(z)}\\
    \end{aligned}
\end{equation}
Lastly, we relate
$\mdmahalos=f_{e,\rm halos} \, \times \, \mdmacosmic$. The dashed lines in Fig.~\ref{fig:dm_z_sdss} represent \dmahalos,~and we note that the \dmhalos \, for the FRB sightline is well below this value at all redshifts. 

There are two major sources of uncertainty in estimating \dmhalos. 
First, stellar masses are obtained from SED fitting and have uncertainties of the order of 0.1 dex. In terms of halo masses, this translates to an uncertainty of $\sim0.15$ dex if the mean SHMR is used. 
Second, there is scatter in the SHMR which is also a function of the stellar mass. 
Note that the intervening halos have stellar masses $\sim 10^{10.6}~\mmsun$. 
This corresponds to an uncertainty in the halo mass of $\sim0.25$ dex \citep{moster+13}. In Fig.~\ref{fig:dm_z_sdss}, we have varied stellar masses by 0.1 dex and have depicted the variation in \dmhalos~through the shaded regions. If instead, we varied the stellar masses by 0.16 dex, thus mimicking a variation in halo masses by nearly 0.25 dex, the scatter increases by roughly 10 \dmunits~in Fig. \ref{fig:dm_z_sdss}a. and about 20 \dmunits~in Fig. \ref{fig:dm_z_sdss}b at $z = \zfrb$.

For the remainder of our analysis, we shall use the estimate for \dmhalos~corresponding to \rmax~= 1, i.e. \dmhalos = 12 \dmunits~and is bounded between 7 \dmunits~and 28 \dmunits, while bearing in mind that it may be roughly two times larger if the radial extent of halo gas exceeds \rvir. For the galaxies with photometric redshifts only, we shall adopt $z_{\rm phot}$ and thus estimate no contribution to \dmhalos.

\subsection{\dmigm\ and \dmcosmic}

We now proceed to estimate the other component of \dmcosmic, \dmigm, the contribution from diffuse gas outside halos. In this section, we discuss two approaches to estimating \dmigm.

\begin{enumerate}
    \item The diffuse IGM is assumed to be uniform and isotropic. This implies its DM contribution is completely determined by cosmology and our assumptions for \dmhalos. This is equivalent to estimating the cosmic average of the IGM contribution, \dmaigm.
    
    \item  Owing to structure in the cosmic web, 
    the IGM is not assumed to be uniform.
    We infer the 3D distribution of the cosmic web 
    using the galaxy distribution and
    then use this to compute \dmigm.
\end{enumerate}

We consider each of these in turn.

\begin{figure}[th!]
    \includegraphics[width=\linewidth]{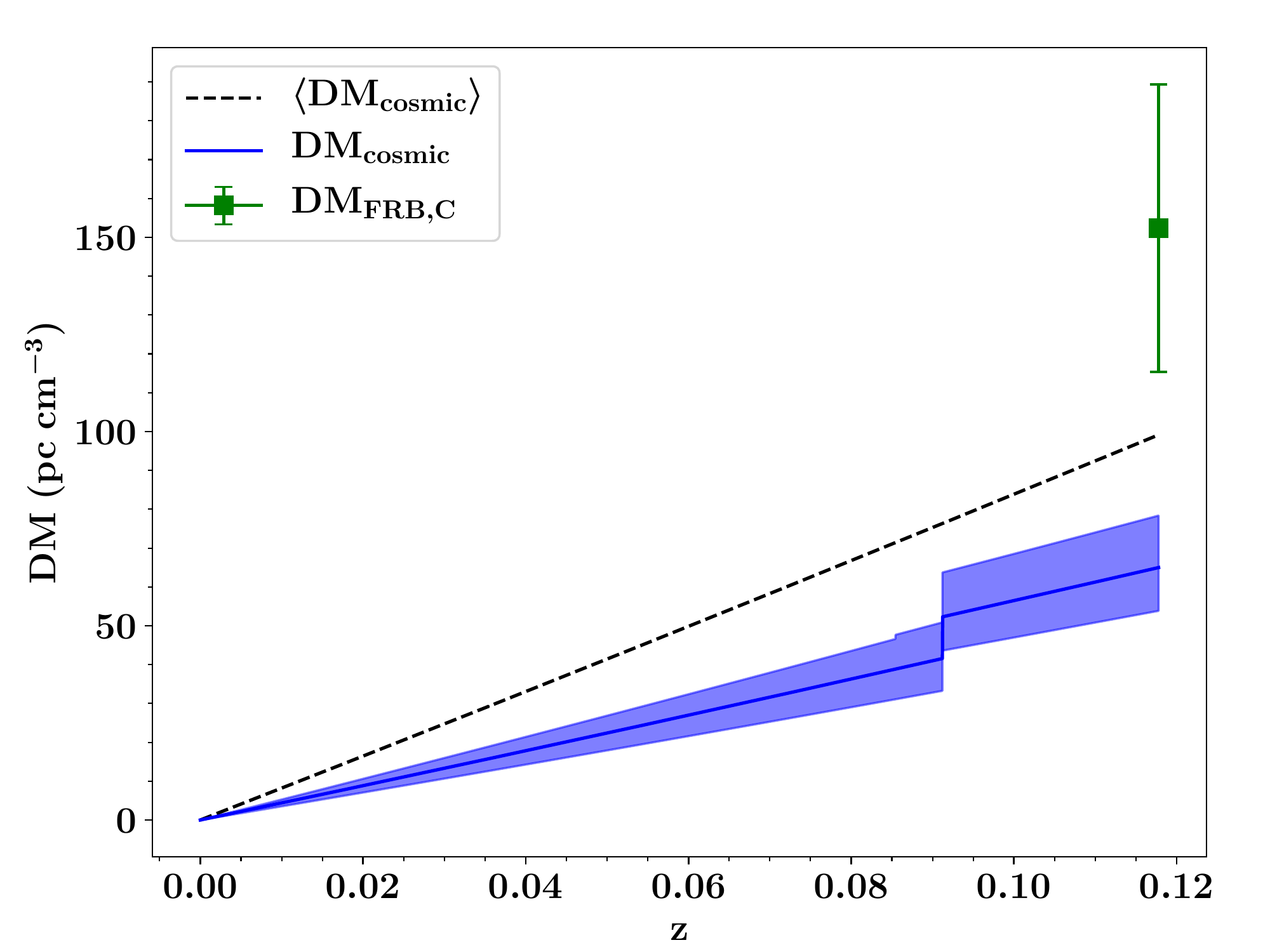}
    \caption{\footnotesize\textbf{ \dmcosmic~vs redshift}. The solid blue line corresponds to  $\mdmcosmic=\mdmhalos+\mdmaigm$ with \fhot~= 0.75 and  \rmax~= 1. 
    The shaded region represents the quadrature sum of uncertainties in \dmhalos\ (allowing for 0.1~dex variation in stellar mass) and the IGM (taken to be 20\% of \dmigm). The green point is \dmfrbc\ (i.e. $\mdmfrb-\rm DM_{MW}-\mdmhost$). The errorbars correspond to the uncertainty in $\mdmhost$, which is $37~\mdmunits$. The black line represents \dmacosmic.}
    \label{fig:dm_cosmic}
\end{figure}
\begin{figure}[th!]
    \includegraphics[width=\linewidth]{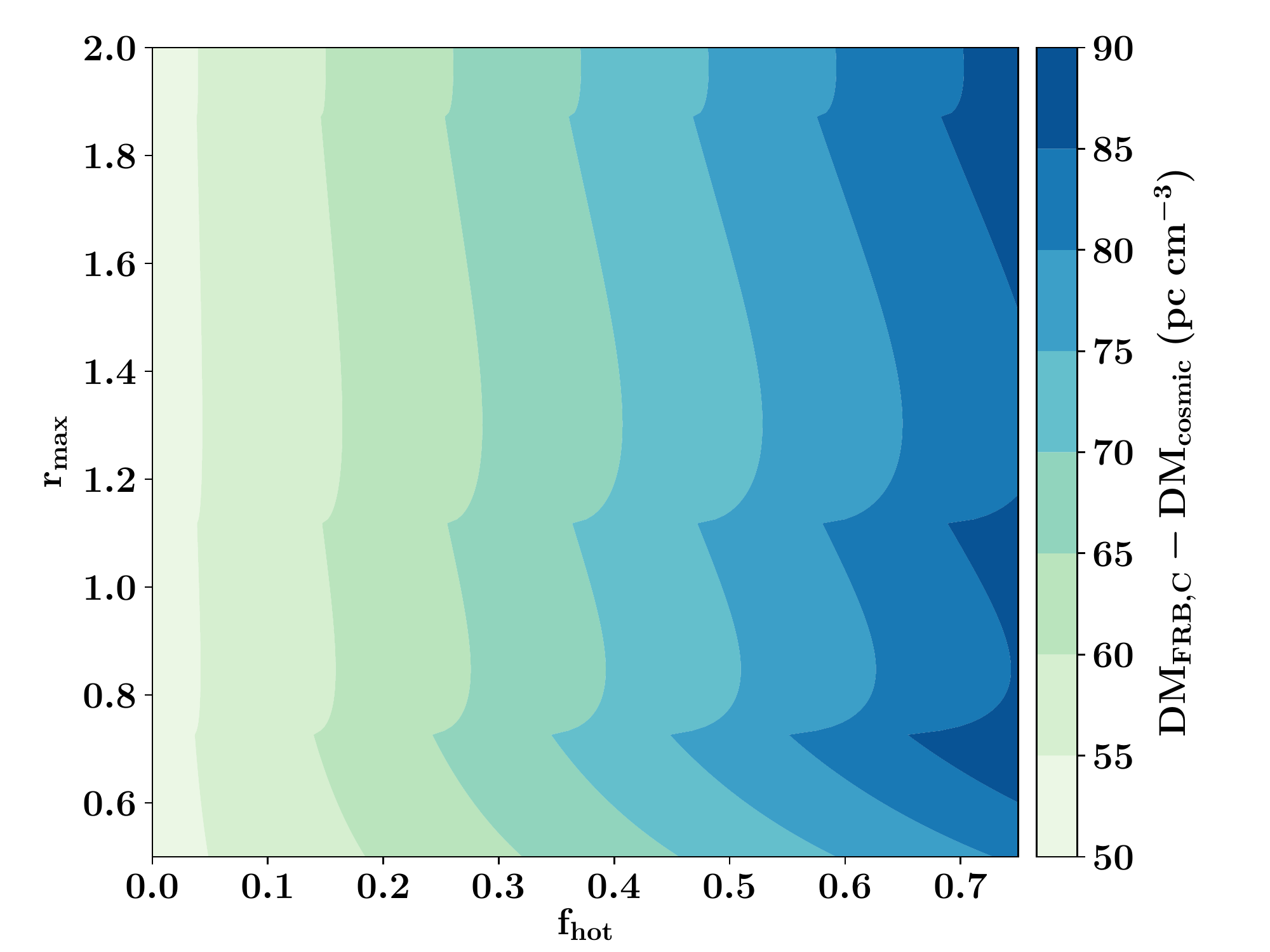}
    %    {\bf Figure \figfrb:} 
    \caption{\footnotesize\textbf{ \dmcosmic~compared to \dmfrbc~as a function of halo model parameters}. Here, \dmcosmic~is defined as \dmhalos+\dmaigm~and depends on two key parameters, \fhot\ and \rmax. \fhot\ is the fraction of baryonic matter present as hot gas in halos, and \rmax~ is the radial extent in units of \rvir\ up to which baryons are present in the halo. At low \fhot\ and \rmax~values, \dmhalos~is small and $\mdmcosmic\approx\mdmaigm\approx\mdmacosmic$. Towards higher \fhot~and \rmax~values, \dmaigm~decreases and \dmhalos~increases. However, $\mdmhalos<\mdmahalos$. Thus \dmcosmic~decreases further compared to \dmfrbc. In summary, \dmcosmic~estimated this way being small is a reflection of the lower than average contribution from \dmhalos.}
    \label{fig:dm_cosmic_contourf}
\end{figure}

\begin{figure*}[t!]
    \centering
    \includegraphics[width=\linewidth]{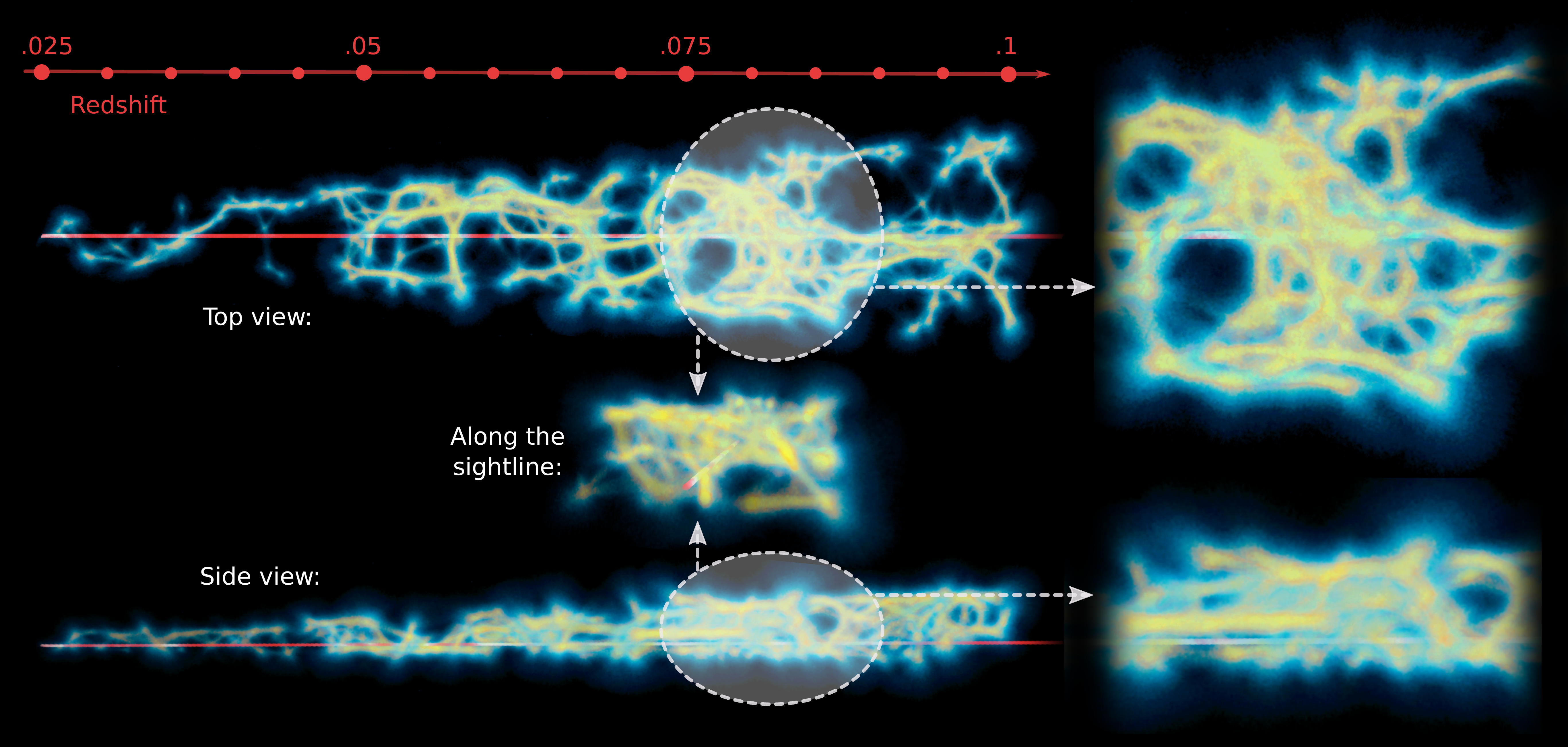}
    \caption{\footnotesize  \textbf{A 3D model of the cosmic web in physical coordinates reconstructed using the MCPM} \textit{Left, top}: The red line passing through the web represents the FRB sightline where light is assumed to travel from right to left. The cosmic web reconstruction \protect{\citep{polyphorm}} is shown color-coded by the steady-state Physarum particle trace density (yellow being high and black being low). The red line with ticks along the top shows the horizontal scale of the reconstruction in redshift.  In the vertical direction, the reconstructed region of the web spans an angular diameter of $800'$ on the sky. \textit{Left, bottom}: A rotated view of the reconstruction. The FRB sightline falls within a narrow strip of the SDSS footprint, and the vertical size in the side view is smaller than that in the top view. \textit{Left, center}: A view along the sightline (which is again visible in red) of a high-density region enclosed by the translucent circles in the top and side views. \textit{Right}: Two close-up views of the locations indicated by the circles on the left. 
    }
    \label{fig:slime}
\end{figure*}

\subsubsection{\dmaigm}

Approach 1 is an approximation of \dmigm. We define:
\begin{equation}
    \mdmaigm = \mdmacosmic - \mdmahalos
\label{eqn:avgdmhalos}
\end{equation}
Naturally, \dmaigm~is redshift dependent and depends on our parameterization of \dmahalos, 
i.e., on \fhot\ and \rmax.
At $z = \mzzfrb$ for \fhot=0.75 and \rmax=1, $\mdmaigm\approx\vdmaigm~\mdmunits$, i.e. about 54\% of \dmacosmic. 

Adopting this value of \dmaigm\, we can estimate \dmcosmic\ towards FRB~190608 by 
combining it with our estimate of \dmhalos\ 
(Fig.~\ref{fig:dm_cumul}).
This is presented as the blue, shaded curve in 
Fig. \ref{fig:dm_cosmic} using our fiducial estimate for \dmhalos\ ($\mfhot=0.75$, $\mrmax=1$). 
This \dmcosmic\ estimate is roughly $90~\mdmunits$ less
than \dmfrbc, and the discrepancy would be larger
if one adopted a smaller \dmmwhalo\ value than 40\,\dmunits\ \citep[e.g.][]{keating_pen2020}.
We have also computed \dmcosmic~for different combinations of \fhot~and \rmax~and show the results in 
Fig.~\ref{fig:dm_cosmic_contourf}.

First, we note that the \dmcosmic~estimate is always lower than \dmfrbc. Second, it is not intuitive that the estimate is closer to \dmfrbc~when $\mfhot\approx0$ (i.e., $\mdmhalos\approx0$). This arises from our definition of \dmaigm, i.e.\ $\mfhot=0$ implies $\mdmahalos=0$ or $\mdmaigm=\mdmacosmic$. As $\mdmacosmic=100~\mdmunits$ is independent of \fhot~and \rmax, the estimate is close to \dmfrbc. For all higher \fhot, \dmaigm~is smaller and \dmhalos~is insufficient to add up to \dmfrbc. In summary, \dmhalos~is consistently lower than \dmahalos~for the parameter range we explored. This results in the \dmcosmic~thus estimated being systematically lower than \dmfrbc.

%%%%%%%%%%%%%%%%%%%%%%%%%%%%%%%%%%%%%%%%%%%%%%%%%%%%%%%%%%%%%%%%%%%%%%
\subsubsection{Cosmic web reconstruction}
% Joe Burchett
\label{sec:cosmicWeb}

As described in Sec. \ref{sec:fgHalos}, 
the localization of FRB~190608 to a region with SDSS coverage enables modeling of the DM contribution from individual halos along the line of sight. 
It also invites the opportunity to consider cosmic gas residing within the underlying, large-scale structure. %i.e. the cosmic web.  
Theoretical models predict shock-heated gas within the cosmic web as a natural consequence of structure formation \citep{Cen:1999yq, Dave+2001}, and indeed, FRBs offer one of the most promising paths forward in detecting this elusive material \citep{jpm+2020}.  

Using the SDSS galaxy distribution within $400'$ of the FRB sightline, we employed the Monte Carlo Physarum Machine (MCPM) cosmic web reconstruction methodology introduced by 
\citet{Burchett:2020_slime} to map the large-scale structure intercepted by the FRB sightline.  Briefly, the slime mold-inspired MCPM algorithm finds optimized network pathways between galaxies (analogous to food sources for the Physarum slime mold) in a statistical sense to predict the putative filaments in which they reside.  The galaxies themselves occupy points in a three-dimensional (3D) space determined by their sky coordinates and the luminosity distances indicated by their redshifts.  At each galaxy location, a simulated chemo-attractant weighted by the galaxy mass is emitted at every time step.  Released into the volume are millions of simulated slime mold `agents', which move at each time step in directions preferentially toward the emitted attractants.  Thus, the agents eventually reach an equilibrium pathway network producing a connected 3D structure representing the putative filaments of the cosmic web.  The trajectories of the agents are aggregated over hundreds of time steps to yield a `trace', which in turn acts as a proxy for the local density at each point in the volume (see \citealt{Burchett:2020_slime} for further details).

   \begin{figure}[h!]
     \includegraphics[width=\linewidth]{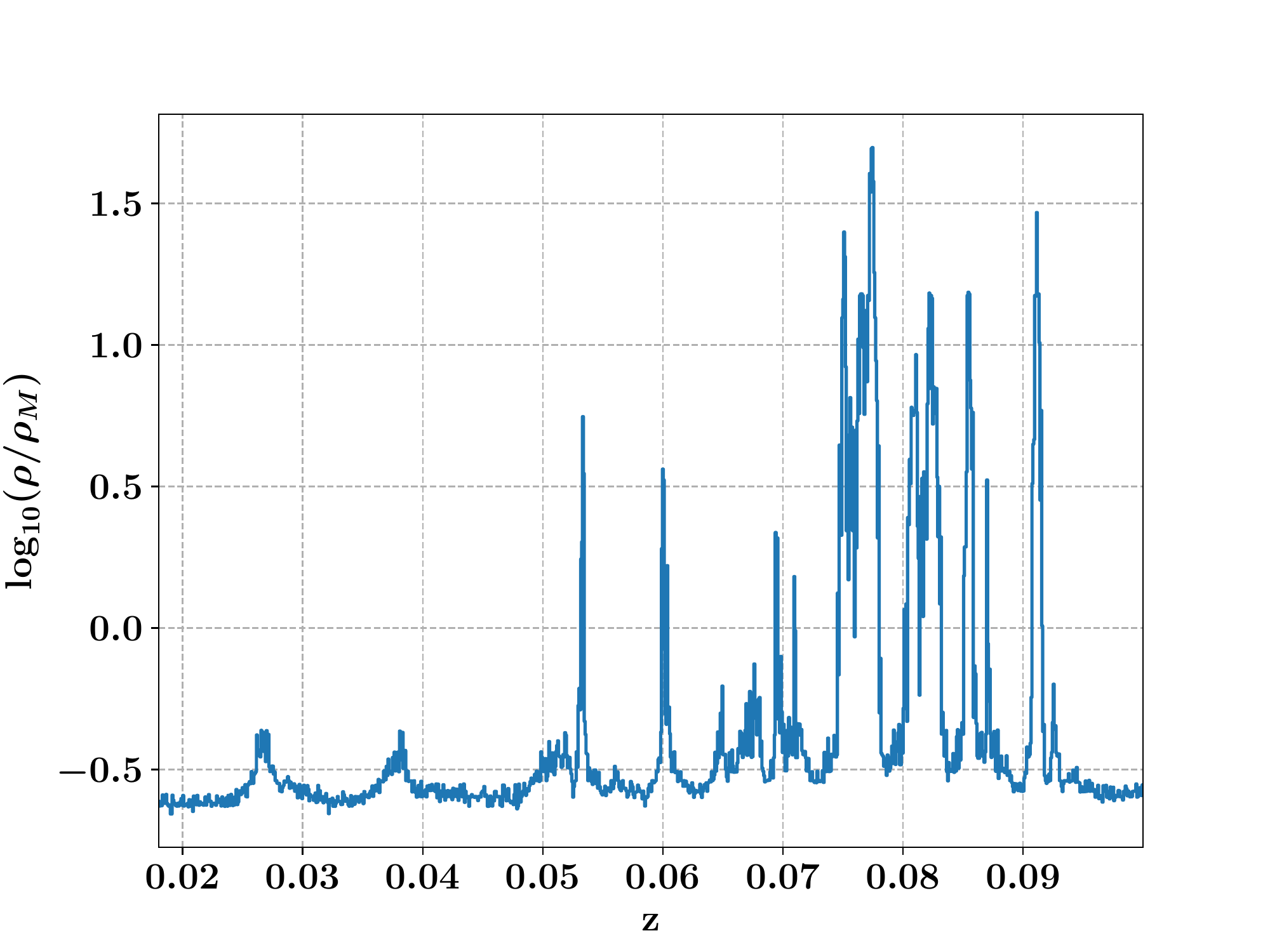}
     \caption{\footnotesize \textbf{Cosmic web density estimate from MCPM.} We show the MCPM-derived cosmic overdensity as a function of redshift along the line of sight to FRB 190608.  We first produced our cosmic web reconstruction from SDSS galaxies within 400 arcmin of the sightline and then calibrated the MCPM trace (see text) with the cosmic matter density from the Bolshoi-Planck simulation.  Note that there are apparently no galaxy halos ($\rm \rho>100\rho_m$) captured here, although several density peaks arise from large-scale structure filaments. We in turn use the 3D map from MCPM to model the diffuse IGM gas and produce \dmigm~estimates.}
     \label{fig:rho_slime}
 \end{figure}

Our reconstruction of the structure intercepted by our FRB sightline is visualized in Fig.~\ref{fig:slime}. The MCPM methodology simultaneously offers the features of 1) producing a continuous 3D density field defined even relatively far away from galaxies on Mpc scales and 2) tracing anisotropic filamentary structures on both large and small scales.

 \begin{figure*}
    \includegraphics[width=\textwidth]{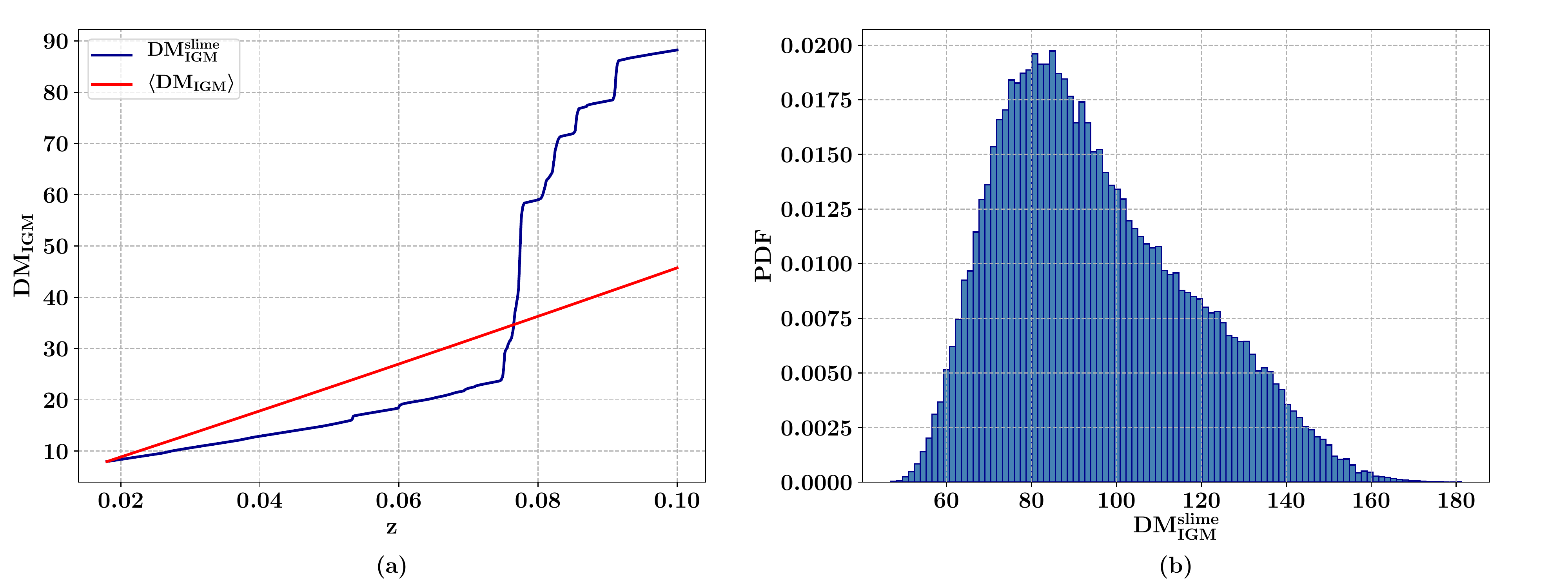}
    \caption{\footnotesize \textbf{\dmslime \, from MCPM density estimate.} \textit{(Left)} A comparison of \dmigm \, obtained from the MCPM analysis (blue) and \dmaigm \, (red) assuming \fhot~= 0.75 and \rmax~= 1. Below $\rm z = 0.018$, where the MCPM density estimate is not available, \dmslime~is assumed to be equal to \dmaigm. At $\rm z = 0.1$, \dmslime\ is nearly twice \dmaigm. \textit{(Right)} The \dmslime\ PDF estimated from accounting for the uncertainties in the Bolshoi-Planck mapping from particle trace densities to physical overdensities. The full-width at half-maximum (FWHM) of each density peak is independently varied by a factor within 0.5 dex and a cumulative DM is computed. This estimate of the PDF is obtained from 100,000 realizations of \dmslime. \dmslime = 88 \dmunits\ for $z\leq$ 0.1, and its distribution is asymmetric with a standard deviation of $\sim15~\mdmunits$.}
    \label{fig:slime_DM}
 \end{figure*}

With the localization of FRB 190608 both in redshift and projected sky coordinates, we retrieved the local density as a function of redshift along the FRB sightline from the MCPM-fitted volume. The SDSS survey is approximately complete to galaxies with \mstar\ $\geq 10^{10.0} \mmsun$, which translates via abundance matching \citep{moster+13} to \mhalo\ $\geq 10^{11.5} \mmsun$. Therefore, we only used galaxies and halos above these respective mass limits in our MCPM fits for the SDSS and Bolshoi-Planck datasets. This prevents us from extending the redshift range of our analysis beyond 0.1, as going further would require a higher mass cutoff and therefore a much sparser sample of galaxies on which to perform the analysis. On the lower end of the redshift scale, there are fewer galaxies more massive than $10^{10.0} \mmsun$ (see Fig. \ref{fig:rho_z}) and therefore the MCPM fits are limited to $z>0.018$. To translate the MCPM density metric \rhophys\ 
 to a physical overdensity $\delta \rho / \rho_m$, we applied MCPM to the dark matter-only Bolshoi cosmological
simulation, where the matter density $\rho_m$ is known at each point.  Rather than galaxies, we fed the MCPM locations and masses of dark matter halos \citep{Behroozi:2013aa}.   We then calibrated \rhophys\ to $\rho / \rho_m$ as detailed by \citet{Burchett:2020_slime}.  This produces a mapping to physical overdensity, albeit less tightly constrained than that of \citet{Burchett:2020_slime} due to the sparser dataset we employ here.  For densities $\rho \gtrsim \rho_m$, we estimate a roughly order of magnitude uncertainty in $\rho/ \rho_m$ derived along the line of sight. Fig.~\ref{fig:rho_slime} shows the density relative to the average matter density as a function of redshift.
 
 The electron number density $n_e(z)$ is obtained by multiplying $\bar n_e(z)$ from equation \ref{eqn:DMzRelation} with the MCPM estimate for $\rm \rho/\rho_m$. 
 Last, we integrate $n_e$ to 
 estimate \dmslime~and recover
 $\mdmslime = 78~\mdmunits$
 for the redshift interval $z=[0.018,0.1]$ (see Fig.~\ref{fig:slime_DM}a). \dmslime~is nearly double the value of \dmaigm~at $z=0.1$~assuming \fhot~= 0.75 and \rmax~= 1.

 The Bolshoi-Planck mapping from the trace densities to physical overdensity includes an uncertainty of $\sim0.5$ dex in each trace density bin. To estimate the uncertainty in \dmslime, we first identify the peaks in Fig.~\ref{fig:rho_slime}. For all pixels within the full-width at half-maximum (FWHM) of each peak, we vary the relative density by a factor that does not exceed 0.5 dex. This factor is drawn from a uniform distribution in log space. Each peak was assumed to be independent and thus varied by different factors, and \dmslime\ was recomputed. From 100,000 such realizations of \dmslime, we estimated a probability density function (PDF) (Fig.~\ref{fig:slime_DM}b). The 25th and 75th percentiles of this distribution are 75 \dmunits~ and 110 \dmunits, respectively and the median value is 91 \dmunits. For the redshift intervals excluded, we assume $n_e = \bar n_e$ and estimate an additional 16 \dmunits~to \dmigm\ (8 \dmunits~for $z<0.018$ and 8~\dmunits~for $z>0.1$), increasing \dmigm~to 94 \dmunits. This is justified by comparing Fig.~\ref{fig:rho_z} and Fig.~\ref{fig:rho_slime} to assess 
 that there are no excluded overdensities that can contribute more than a few \dmunits~over the average value. In conclusion, we estimate \dmigm~= 94 \dmunits~with the 25th and 75th percentile bounds being 91 \dmunits~and 126 \dmunits.
 
With detailed knowledge of the IGM matter density, one can consider defining the boundary of a halo more precisely. A natural definition for the halo radius would be where the halo gas density and the IGM density are identical. Therefore, we tested whether the \rmax~obtained would significantly differ from the chosen value of unity, and thus produce substantially different \dmhalos, for the intervening halos. We estimated \rmax~using this condition by setting the IGM density as the value obtained from the MCPM model at each halo redshift, yielding $\mrmax\approx 1.3-2.2$ for the halos. \dmhalos~estimated using these \rmax~values for the halos is $~\approx30~\mdmunits$ as only the first two halos in Table \ref{tab:fg_gals} contribute. This is only slightly higher than the upper bound obtained previously for $\mrmax=1$ and therefore, we we choose to continue with the \dmhalos~value initially estimated using $\mrmax = 1$.
 Finally, our cosmic web reconstruction from the MCPM algorithm also allows us to refine our estimate of expected intervening galaxy halos in the KCWI FoV, $\langle n^{\rm KCWI}_{\rm halos}\rangle = 0.23$, presented in Section~\ref{sec:dataset}. Given the inferred overdensity as a function of redshift along the line of sight, $\rho/\rho_m(z)$, and the co-moving volume element given by the KCWI FoV as a function of redshift, $dV(z)$, we can then just scale $\langle n^{\rm KCWI}_{\rm halos}\rangle$ by $\alpha \equiv \frac{ \int \rho/\rho_m(z) dV(z)\,dz}{\int dV(z)\,dz}$. In our case, we have obtained $\alpha = 1.66$, and then our refined $\langle n^{\rm KCWI}_{\rm halos}\rangle = 0.38$. This number is still small and, thus, fully consistent with a lack of intervening halos found in the KCWI FoV.

%%%%%%%%%%%%%%%%%%%%%%%%%%%%%%%%%%%%%%%%%%%%%%%%%%%%%%%%%%%%%%%%%%%%%%%%%%%%%%%%%%%%%%%%%%%%%%%%%
%%%%%%%%%%%%%%%%%%%%%%%%%%%%%%%%%%%%%%%%%%%%%%%%%%%%%%%%%%%%%%%%%%%%%%%%%%%%%%%%%%%%%%%%%%%%%%%%%
\section{Cosmic contributions to the Rotation Measure and Temporal Broadening}
\label{sec:RM}

We briefly consider the potential contributions of foreground
galaxies to FRB~190608's 
observed temporal broadening and rotation measure.
As evident in Table~\ref{tab:fg_gals},
there is only a single halo within 200\,kpc of 
the sightline with $z \le \mzzfrb$.  It has 
redshift $z=0.09122$ and an estimated halo
mass $\mmhalo = 10^{12} \mmsun$.

FRB 190608 exhibits a large, 
frequency-dependent pulse width 
%($\tau\sim\nu^{-4}$). 
$\tau = 3.3 \, \rm~ms$ at 1.28~GHz \citep{day+2020}, which
exceeds the majority of previously reported 
pulse widths \citep{frbcat}. 
Pulses are broadened when interacting with turbulent media. While we expect a scattering pulse width much smaller than a few milliseconds from the diffuse IGM alone \citep{Macquart2013}, we consider the possibility that the denser halo gas at $z = 0.09122$ contributes significantly to FRB 190608's intrinsic pulse profile. 
Here, we estimate the extent of such an effect, emphasizing that the geometric
dependence of scattering greatly favors gas in intervening halos as opposed to the host galaxy.

Assuming the density profile as described 
in Section~\ref{sec:fgHalos}
(extending to \rmax=1), the maximum electron density ascribed to the halo is at its impact parameter 
$b = 158$\,kpc: $n_e\sim10^{-4} \, \rm cm^{-3}$. Note that $b$ is much greater than the impact parameter of the foreground galaxy of FRB 181112 \citep[29~kpc,][]{frb181112} and indeed that of the host or the Milky Way with FRB 190608's sightline. The entire intervening
halo can be thought of effectively as a ``screen" whose thickness is the length the FRB sightline intersects with the halo, $\Delta L = 265~\rm kpc$. We assume the turbulence is described by a Kolmogorov distribution of density fluctuations with an outer scale $L_0=1 \rm~pc$. This choice of $L_0$ arises from assuming stellar activity is the primary driving mechanism. To get an upper bound on the pulse width produced, we also assume the electron density is equal to $10^{-4}\rm ~cm^{-3}$ for the entire length of the intersected sightline. Following the scaling relation in equation 1 from \citealt{frb181112}, we obtain: 
\begin{equation}
    \begin{aligned}
        \tau_{\rm1.4~GHz}<~&0.028~\rm ms~
        \alpha^{12/5}\left(\frac{n_e}{10^{-4}~\rm cm^{-3}}\right)^{12/5}\\
        &\times\left(\frac{\Delta L}{265~\rm kpc}\right)^{6/5}\left(\frac{L_0}{1~\rm pc}\right)^{-4/5}
    \end{aligned}
\end{equation}
Here, $\alpha$ is a dimensionless number that encapsulates the root mean-squared amplitude of the density fluctuations and the volume-filling fraction of the turbulence. It is typically of order unity. We note that our chosen value of $L_0$ presents an upper limit on the scattering timescale. Were $L_0\gg 1~\rm pc$ (e.g. if driven by AGN jets), $\tau\ll 0.03~\rm ms$. The observed scattering timescale exceeds our conservative upper bound by two orders of magnitude. One would require $n_e>6\times10^{-4}~\rm cm^{-3}$ to produce the observed pulse width. 
This exceeds the maximum density estimation through the halo, even for the relatively
flat and high \fhot\ assumed.
We thus conclude that the pulse broadening for FRB~190608 
is not dominated by intervening halo gas.

FRB 190608 also has a large estimated ${\rm RM}_{\rm FRB} = 353\pm 2$\,\rmunits~\citep{day+2020}. 
We may estimate the RM contributed by the intervening
halo, under the assumption that 
its magnetic field is characterized by the 
equipartition strength magnetic fields in galaxies ($\sim10~\mu G$) \citep{Basu_Roy13}.  We note that this exceeds
the upper limit imposed on gas in the halo intervening
FRB~181112 \citep{frb181112}.

We estimate:
\begin{equation}
    \begin{aligned}
        \rm RM_{halos} =  0.14~\rm rad~m^{-2}&\left(\frac{\rm B_{\parallel}}{10~\rm \mu G}\right)\left(\frac{\Delta L}{265~\rm kpc}\right)\\
        &\times\left(\frac{n_e}{10^{-4}~\mathrm{cm}^{-3}}\right)
    \end{aligned}
\end{equation}
and conclude that
it is highly unlikely that the RM 
contribution from intervening halos 
dominates the observed quantity.

\section{Concluding Remarks}

\begin{table*}
\caption{\footnotesize  Contributions to \dmfrb\ from foreground components\label{tab:dm_summary}}
\begin{center}
    \begin{tabular}{|c|c|c|c|l|}
        \hline
        Component & Sub component & Notation & Value (\dmunits)  & Comments \\
        \hline
        Host Galaxy & ISM & $\rm DM_{host, ISM}$ & 47-117 &  From  \citet{chittidi+20}\\
        \cline{2-5}
                    & Halo & $\rm DM_{host, halo}$ & 15-41 &  From \citet{chittidi+20}\\
        \hline
        Foreground cosmos & Intervening halos & \dmhalos & 7-28\tablenotemark{a} &  Using SDSS spectroscopic galaxies\\
                    & & & &  and 0.16 dex scatter in M*\\
        \cline{3-5}
                    & & \dmahalos & 45\tablenotemark{a} &  Average assuming the Aemulus HMF and \\
                    & & & & Planck 15 cosmology\\
        \cline{2-5}
                    & Diffuse IGM & \dmigm & 91-126 &  25th and 75th percentiles using\\
                    & & & & the MCPM method \\
        \cline{3-5}
                    & & \dmaigm & \vdmaigm\tablenotemark{a} &  Average assuming the Aemulus HMF and  \\
                    & & & & Planck 15 cosmology\\
        \hline
        Milky Way & ISM & $\rm DM_{MW, ISM}$ & 38 &  From \citet{ne2001}\\
        \cline{2-5}
         & Halo & $\rm DM_{MW, halo}$ & 40 &  From \citet{xyz19}\\
        \hline
    \end{tabular}
\end{center}
\tablenotetext{a}{\footnotesize Assuming \fhot~= 0.75 and \rmax~= 1}
\end{table*}

To summarize, we have created a semi-empirical model of the matter distribution in the foreground universe of FRB 190608 using spectroscopic and photometric data from the SDSS database and our own KCWI observations. We modeled the virialized gas in intervening halos using a modified NFW profile and used the MCPM approach to estimate the ionized gas density in the IGM. Table \ref{tab:dm_summary} summarizes the estimated DM contributions from each of the individual foreground components. Adding \dmahalos~and \dmigm~for this sightline, we infer $\mdmcosmic = 98-154 \mdmunits$, which is comparable to \dmacosmic = 100 \dmunits.
The majority of \dmcosmic~is accounted for by the diffuse IGM, implying that most of the ionized matter along this sightline is not in virialized halos. We found only 4 galactic halos within 550 kpc of the FRB sightline and only 1 halo within 200 kpc. We found no foreground object in emission from our $\sim 1$ sq. arcmin KCWI coverage and no galaxy group or cluster having an impact parameter of less than its virial radius with our FRB sightline.

We also find it implausible that the foreground structures are dense enough to account for either the pulse broadening or the large rotation measure of the FRB. We expect the progenitor environment and the host galaxy together are the likely origins of both Faraday rotation and turbulent scattering of the pulse \citep[discussed in further detail by][]{chittidi+20}.

The results presented here are not the first attempt to measure \dmcosmic~along FRB sightlines by accounting for density structures. \citet{Li+19} estimated \dmcosmic~(termed \dmigm~in their paper) for five FRB sightlines, making use of the 2MASS Redshift Survey group catalog \citep{Lim+17} to infer the matter density field along their lines of sight. They assumed NFW profiles around each identified group. This enabled them to estimate the DM contribution from intervening matter for low DM ($\mdmcosmic+\mdmhost <100 ~\mdmunits$) FRBs. Our approach differs in the methods used to estimate \dmcosmic. The precise localization of FRB 190608 allows us to estimate \dmhalos~and \dmigm~separately.  \citet{Li+19} were limited by the large uncertainties ($\sim10'$) in the FRB position and therefore their estimates of \dmacosmic~depended on the assumed host galaxy within the localization regions. Furthermore, the MCPM model estimates the cosmic density field, and thus the ionized gas density of the IGM, due to filamentary large-scale structure.  We note that our estimates of $n_e$ from the MCPM model in overdense regions is similar to their reported values ($10^{-6}-10^{-5}$\,~cm$^{-3}$). This naturally implies our \dmcosmic~estimates are of the same order of magnitude around $z=0.1$ as their estimate. Together with the results presented by \citet{chittidi+20}, our study represents a first of its kind: an observationally driven, detailed DM budgeting along a well-localized FRB sightline. We have presented a framework for using FRBs as quantitative probes of foreground ionized matter. Although aspects of this framework carry large uncertainties at this juncture, the methodology should become increasingly precise as this nascent field of study matures. For instance, our analysis required spectroscopic data across a wide area (i.e. a few square degrees) around the FRB, which enabled us to constrain the individual contributions of halos and also to model the cosmic structure of the foreground IGM. An increase in sky coverage and depth of spectroscopic surveys would enable the use of cosmic web mapping tools like the MCPM estimator with higher precision and on more FRB sightlines. Upcoming spectroscopic instruments such as DESI and 4MOST will map out cosmic structure in greater detail and will, no doubt, aid in the use of FRBs as cosmological probes of matter.

We expect FRBs to be localized more frequently in the future, thanks to thanks to continued improvements in high-time resolution backends and real-time detection systems for radio interferometers. One can turn the analysis around and use the larger set of localized FRBs to constrain models of the cosmic web in a region and possibly perform tomographic reconstructions of filamentary structure. Alternatively, by accounting for the DM contributions of galactic halos and diffuse gas, one may constrain the density and ionization state of matter present in intervening galactic clusters or groups. Understanding the cosmic contribution to the FRB dispersion measures can also help constrain progenitor theories by setting upper limits on the amount of dispersion measure arising from the region within a few parsecs of the FRB. We are at the brink of a new era of cosmology with new discoveries and constraints coming from FRBs.

\textit{Acknowledgments:}
Authors S.S., J.X.P.,
N.T., J.S.C. and R.A.J., as members of the Fast and Fortunate for FRB
Follow-up team, acknowledge support from 
NSF grants AST-1911140 and AST-1910471.
J.N.B. is supported by NASA through grant number HST-AR15009 from the Space Telescope Science Institute, which is operated by AURA, Inc., under NASA contract NAS5-26555.
This work is supported by the Nantucket Maria Mitchell Association. R.A.J. and J.S.C. gratefully acknowledge the support of the Theodore Dunham, Jr. Grant of the Fund for Astrophysical Research.
K.W.B., J.P.M, and R.M.S. acknowledge Australian Research Council (ARC) grant DP180100857.
 A.T.D. is the recipient of an ARC Future Fellowship (FT150100415).
R.M.S. is the recipient of an ARC Future Fellowship (FT190100155)
 N.T. acknowledges support by FONDECYT grant 11191217.
The Australian Square Kilometre Array Pathfinder is part of the Australia Telescope National Facility which is managed by CSIRO. 
Operation of ASKAP is funded by the Australian Government with support from the National Collaborative Research Infrastructure Strategy. ASKAP uses the resources of the Pawsey Supercomputing Centre. Establishment of ASKAP, the Murchison Radio-astronomy Observatory and the Pawsey Supercomputing Centre are initiatives of the Australian Government, with support from the Government of Western Australia and the Science and Industry Endowment Fund. 
We acknowledge the Wajarri Yamatji as the traditional owners of the Murchison Radio-astronomy Observatory site. 
Spectra were obtained at the W. M. Keck Observatory, which is operated as a scientific partnership among Caltech, the University of California, and the National Aeronautics and Space Administration (NASA). The Keck Observatory was made possible by the generous financial support of the W. M. Keck Foundation. The authors recognize and acknowledge the very significant cultural role and reverence that the summit of Mauna Kea has always had within the indigenous Hawaiian community. We are most fortunate to have the opportunity to conduct observations from this mountain.
\software{
KCWIDRP \citep{KCWIDRP},
SEP \citep{Barbary2016,SExtractor},
MARZ \citep{marz},
HMFEmulator \citep{HMF},
CIGALE \citep{cigale},
Astropy \citep{astropy:2018},
Numpy \citep{numpy},
Scipy \citep{scipy},
Matplotlib \citep{Hunter:2007},
Polyphorm \citep{polyphorm}
}

\appendix
\section{Cosmic diffuse gas fraction}
\label{appendix:f_d}

Central to an estimate of \dmcosmic\ is the fraction of baryons that
are diffuse and ionized in the universe $f_d$. We have presented a brief discussion of $f_d$ in previous works \citep{xyz19,jpm+2020} and provide additional details and an update here.

To estimate $f_d(z)$, we work backwards by defining and estimating the cosmic components that do not contribute to \dmcosmic. These are:

\begin{enumerate}
    \item Baryons in stars, $\rho_{\rm stars}$.  This quantity is estimated from galaxy
    surveys and inferences of the stellar initial mass function
    \citep{MadauDickinson14}.
    \item Baryons in stellar remnants and brown dwarfs, $\rho_{\rm remnants}$.
    This quantity was estimated by \cite{Fukugita04} to be $\approx 0.3 \rho_{\rm stars}$ at $z=0$.  We adopt this fraction for all cosmic time.
    \item Baryons in neutral atomic gas, $\rho_{\rm HI}$.  This is estimated
    from 21 cm surveys.
    \item Baryons in molecular gas, $\rho_{\rm H_2}$.  This is estimated from
    CO surveys.
\end{enumerate}
One could also include the small contributions from heavy elements, but we ignore this because it is a value smaller than the uncertainty in the dominant components.

\begin{figure}[!hb]
    \centering
    \includegraphics[width=\linewidth]{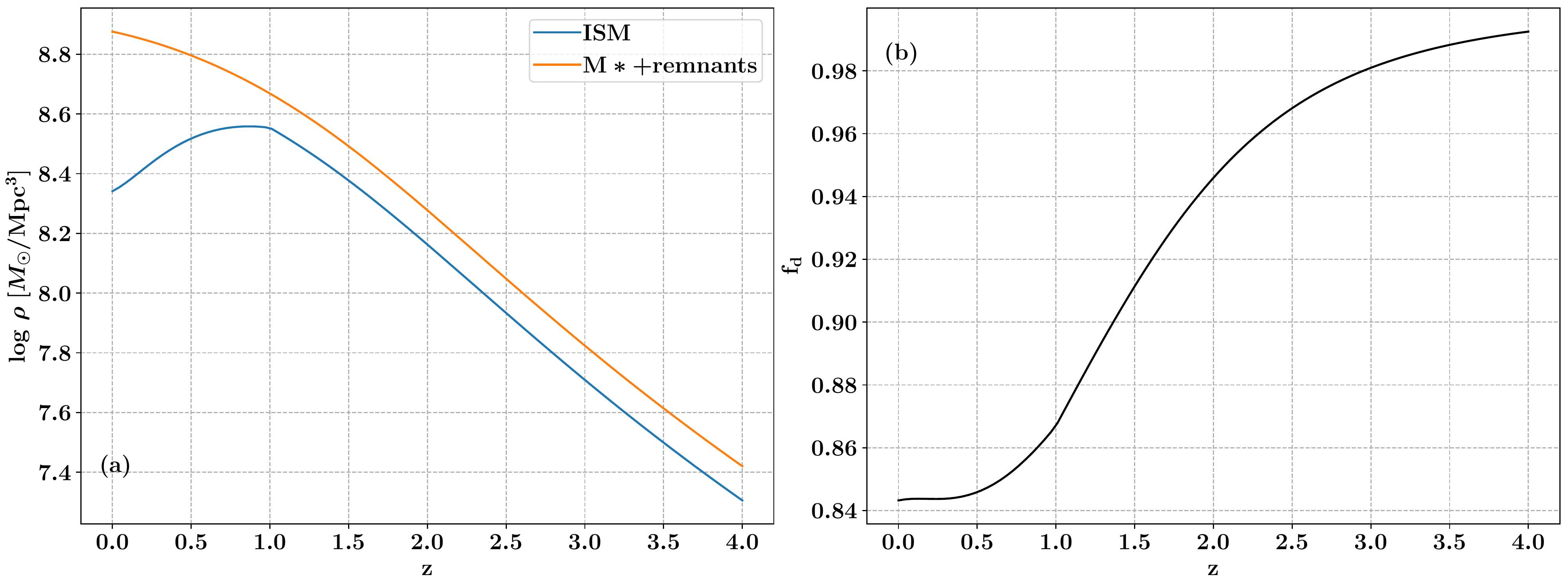}
    \caption{(a) Estimates of the stellar and ISM mass densities
    in galaxies versus redshift.
    (b) Estimate of $f_d$ versus redshift.
    }
    \label{fig:f_d2}
\end{figure}

Altogether, we define 
\begin{equation}
        f_d \equiv 1 - \frac{\rho_{\rm stars}(z) +
                            \rho_{\rm remnants}(z) + 
                            \rho_{\rm ISM}(z)}{\rho_b(z)}
\end{equation}
where we have defined $\rho_{\rm ISM} \equiv \rho_{\rm HI} + \rho_{\rm H_2}$. \cite{Fukugita04} has estimated $\rho_{\rm ISM}/\rho_{\rm stars} \approx 0.38$ at $z=0$
and galaxy researchers assert that this ratio increases to unity by $z=1$ \citep[e.g.][]{tacconi+2020}. For our formulation of $f_d$, we assume $\rho_{\rm ISM}(z)/\rho_{\rm stars}(z)$ increases as a quadratic function with time having values 0.38 and 1 at $z=0$ and 1 respectively, and 0.58 at the half-way time.  The quantity is then taken to be unity at $z>1$. Figure~\ref{fig:f_d2}a shows plots of $\rho_{\rm stars}$ and $\rho_{\rm ISM}$ versus redshift, and Figure~\ref{fig:f_d2}b presents $f_d$. Code that incorporates this formalism is available in the FRB repository\footnote{https://github.com/FRBs/FRB}.

$f_d$, therefore, does not have a simple analytical expression describing it as a function of redshift. One can always approximate $f_d$ as a polynomial expansion in $z$. For $z<1$, one can obtain a reasonable approximation (relative error $<5\%$) by truncating up to the fourth order in $z$:
\textbf{\begin{equation}
    f_d(z) \approx 0.843 + 0.007z - 0.046z^2 + 0.106z^3 - 0.043z^4
\end{equation}}
\bibliography{190608_fg_refs.bib}

\end{document}